\documentclass[preprint,amsmath,showpacs]{revtex4}
\def\DESepsf(#1 width #2){\epsfxsize=#2 \epsfbox{#1}}
\usepackage{graphicx}% Include figure files
\usepackage{color}
\usepackage{amsmath}
\def\bmatrix{\left[\begin{array}}
\def\ematrix{\end{array}\right]}

\begin{document}

%%%%
%    Greek Letters
%

\let\a=\alpha      \let\b=\beta       \let\c=\chi        \let\d=\delta
\let\e=\varepsilon \let\f=\varphi     \let\g=\gamma      \let\h=\eta
\let\k=\kappa      \let\l=\lambda     \let\m=\mu
\let\o=\omega      \let\r=\varrho     \let\s=\sigma
\let\t=\tau        \let\th=\vartheta  \let\y=\upsilon    \let\x=\xi
\let\z=\zeta       \let\io=\iota      \let\vp=\varpi     \let\ro=\rho
\let\ph=\phi       \let\ep=\epsilon   \let\te=\theta
\let\n=\nu
\let\D=\Delta   \let\F=\Phi    \let\G=\Gamma  \let\L=\Lambda
\let\O=\Omega   \let\P=\Pi     \let\Ps=\Psi   \let\Si=\Sigma
\let\Th=\Theta  \let\X=\Xi     \let\Y=\Upsilon

%
%%%

%%%
%    Calligraphic letters
%

\def\cA{{\cal A}}                \def\cB{{\cal B}}
\def\cC{{\cal C}}                \def\cD{{\cal D}}
\def\cE{{\cal E}}                \def\cF{{\cal F}}
\def\cG{{\cal G}}                \def\cH{{\cal H}}
\def\cI{{\cal I}}                \def\cJ{{\cal J}}
\def\cK{{\cal K}}                \def\cL{{\cal L}}
\def\cM{{\cal M}}                \def\cN{{\cal N}}
\def\cO{{\cal O}}                \def\cP{{\cal P}}
\def\cQ{{\cal Q}}                \def\cR{{\cal R}}
\def\cS{{\cal S}}                \def\cT{{\cal T}}
\def\cU{{\cal U}}                \def\cV{{\cal V}}
\def\cW{{\cal W}}                \def\cX{{\cal X}}
\def\cY{{\cal Y}}                \def\cZ{{\cal Z}}
%
%%%%

\newcommand{\Ns}{N\hspace{-4.7mm}\not\hspace{2.7mm}}
\newcommand{\qs}{q\hspace{-3.7mm}\not\hspace{3.4mm}}
\newcommand{\ps}{p\hspace{-3.3mm}\not\hspace{1.2mm}}
\newcommand{\ks}{k\hspace{-3.3mm}\not\hspace{1.2mm}}
\newcommand{\des}{\partial\hspace{-4.mm}\not\hspace{2.5mm}}
\newcommand{\desco}{D\hspace{-4mm}\not\hspace{2mm}}
%machienyi
\renewcommand{\figurename}{Fig.}
\newcommand*{\st}[1]{\textbf{ *** Sming: #1 *** }}
\newcommand*{\mkred}[1]{{\color{red}{#1}}}
\newcommand*{\mkblue}[1]{{\color{blue}{#1}}}
\newcommand{\sigsip}{\ensuremath{\sigma^{\rm{SI}}_p}}
\newcommand{\sigsdp}{\ensuremath{\sigma^{\rm{SD}}_p}}
\newcommand{\sigmav}{\ensuremath{\langle \sigma \, v \rangle}}

%%%%

%
\title{\boldmath
Constraining dark matter capture and annihilation cross sections by searching for
neutrino signature from the Earth core
}
\vfill
\author{Fei-Fan Lee$^{1}$}
%\email{fflee@mail.nctu.edu.tw}
\author{Guey-Lin Lin$^{1}$}
%\email{glin@mail.nctu.edu.tw}
\author{Yue-Lin Sming Tsai$^{2}$}
\affiliation{
 $^1$Institute of Physics,
 National Chiao-Tung University,
 Hsinchu 30010, Taiwan
}
\affiliation{
$^2$National Centre for Nuclear Research,
Ho$\dot{z}$a 69, 00-681 Warsaw, Poland}

\date{\today}
%
%\vskip -1cm
%
%\vfill
%
\begin{abstract}
We study the sensitivity of IceCube/DeepCore detector to dark matter annihilations in the Earth core. We focus on annihilation modes 
$\chi\chi\to \nu \bar{\nu}, \, \tau^+ \tau^-, \,  b \bar{b}$, and $W^+W^-$. Both track and cascade events are considered in our analysis.
By fixing the dark matter annihilation cross section $\langle \sigma\upsilon\rangle$ at some nominal values, we study the sensitivity of IceCube/DeepCore detector to dark matter 
spin-independent cross section $\sigma_p^{\rm SI}$ for $m_{\chi}$
ranging from few tens of GeV to 10 TeV. This sensitivity is compared with the existing IceCube 79-string constraint on the same cross section, which was obtained by searching for 
dark matter annihilations in the Sun. We compare this sensitivity to dark matter direct detection results as well, 
in particular the XENON100 (2012) limit and the parameter regions 
preferred by DAMA and CRESST-II experiments.
We also present IceCube/DeepCore sensitivity to $\langle \sigma\upsilon \rangle$ 
as a function of $m_{\chi}$ by fixing $\sigma_p^{\rm SI}$ 
at XENON100 (2012) and XENON1T limits, respectively. This sensitivity is compared with the preferred dark matter parameter range derived from the combined fitting to PAMELA and AMS02 positron fraction data. 
We conclude that the search for dark matter annihilations in the Earth core  
provides competitive constraints on $\sigma_p^{\rm SI}$ and $\langle \sigma\upsilon \rangle$ in the case of  low-mass dark matter. Particularly, the expected constraint on
$\sigsip$ for 5 years of data taking in IceCube/DeepCore is more stringent than the current IceCube 79-string limit mentioned above.

\end{abstract}
%\vskip2pc]
%
\maketitle

\pagestyle{plain}

\section{Introduction}

Evidences for the dark matter (DM) are provided by many astrophysical observations, although the nature of DM is yet to be uncovered. The most popular candidates for DM are weak interacting massive particles (WIMP), which we shall assume in this work. DM can be detected either directly or indirectly with the former observing the nucleus recoil as DM interacts
with the target nuclei in the detector and the latter detecting final state particles resulting from DM
annihilations or decays. The direct detection is possible because that 
the dark matter particles constantly bombard the Earth as the Earth sweeps through the local 
halos. 
As just stated, the direct detection experiments record the nuclei recoil energy of nuclei-WIMPs scattering. 
At present, DAMA \cite{Bernabei:2010mq}, CoGeNT \cite{Aalseth:2010vx}, and CRESST \cite{Bravin:1999fc} have reported the detection of DM signal 
with the DM mass  $m_\chi$ ranging from few GeV to 50 GeV and the spin-independent scattering cross section 
$\sigsip\sim 10^{-4}$ pb. On the other hand, XENON100 \cite{Aprile:2012nq} only 
collects 2 events which are consist with the background. This result  then sets the limit  $\sigsip < 2\times 10^{-9}$ pb for $m_\chi=55$ GeV.     
Interestingly, recent CDMS II result  \cite{Agnese:2013rvf} reports three signal events 
which gives a p-value 0.19\% (less than $4\sigma$). 
%consistent with the background. 
The corresponding best-fit values of DM parameters are $m_\chi=8.6$ GeV and 
$\sigsip\sim 1.9\times 10^{-5}$ pb. 
%Therefore, in this paper, we will discuss our neutrino events in 
%both discovery and exclusion cases.  

DM can also be detected indirectly by measuring the positron signals 
from Milky Way. PAMELA observed a rise of 
the cosmic ray positron fraction for positron energy greater than  
10 GeV \cite{Adriani:2011xv}. This anomalous enhancement are confirmed by 
Fermi-LAT \cite{Ackermann:2010ij} and the recently released AMS02 first result \cite{Aguilar:2013qda}. 
In the recent AMS02 result, this continuous rise on positron fraction is extended up to   
positron energy $\sim$ 350 GeV. Such a spectral behavior makes the DM annihilation explanation of the data difficult 
because it requires a large boost factor for annihilation cross section $\sigmav$ 
provided the thermal equilibrium for DM in the early universe is reached with $\sigmav\sim 3\times 10^{-26}\,\rm{cm}^3 s^{-1}$. 
For example, several groups 
\cite{Cirelli:2008pk,Kopp:2013eka,DeSimone:2013fia,Yuan:2013eja,Cholis:2013psa,Jin:2013nta,Yuan:2013eba} 
fit the updated galactic positron fraction with AMS02 new result included. 
They have found that the favored DM parameter region is located at 
$m_\chi\sim$ few TeV and $10^{-23}\lesssim\sigmav/\rm{cm}^3 s^{-1}\lesssim10^{-21}$ if DM annihilation channel  $\chi\chi\rightarrow\tau^+\tau^-$
is responsible for the positron excess. The favored DM mass range can be lowered to few hundred GeV if nearly pulsar sources are considered together with  
$\chi\chi\rightarrow\tau^+\tau^-$ annihilations \cite{Yuan:2013eja}.  However the favored values for $\sigmav$ do not decrease much in such a combined fitting.
It is important to note that DM annihilations in the galactic halo are constrained by Fermi-LAT gamma ray observations \cite{Fermi:2012}.  
The constraint on  $\langle\sigma(\chi\chi\to \tau^+\tau^-)\upsilon\rangle$ is in fact located in the preferred  DM parameter region resulting from PAMELA and AMS02 measurements 
for the same range of $m_{\chi}$. 

It has been pointed out some time ago that the preferred DM parameter region by PAMELA and Fermi-LAT measurements  can
be examined  through the observation of neutrinos \cite{Spolyar:2009kx,Buckley:2009kw,Mandal:2009yk}(see also
discussions in Refs.~\cite{Covi:2009xn,Erkoca:2010}) by  IceCube
detector augmented with DeepCore array. Indeed IceCube 22 string result on searching for DM annihilations from galactic halo \cite{Abbasi:2011eq} has set 
the upper limit for $\langle\sigma(\chi\chi\to \tau^+\tau^-)\upsilon\rangle$ comparable  to the required annihilation cross section for explaining PAMELA and Fermi-LAT data. 
The IceCube sensitivity on DM signature from the galactic halo is expected to improve with the data from all 86 strings analyzed. The analysis of DeepCore array data will further enhance the sensitivity in the small $m_\chi$ regime~\cite{Lee:2011nt,Lee:2012pz} which is of interest due to direct detection results mentioned above.  

It is interesting to note that the constraints on DM capture cross section and annihilation cross section $\sigmav$ can be obtained from the searching for DM annihilations from the Earth core. 
The detection of DM induced neutrino signature from the Earth core has been discussed previously \cite{Freese:1985qw,Krauss:1985aaa,Gould:1987ir,Lundberg:2004dn}. It has been shown that the chemical 
composition of the Earth core results in several DM annihilation peaks for $m_{\chi}$ ranging from 20 GeV to 60 GeV.  These peaks do not appear for annihilations  inside the Sun.
Furthermore, DM annihilation rate inside the Sun is completely determined by the capture cross section (dominated by spin dependent component proportional to $\sigsdp$ ) while DM annihilation rate in the Earth core depends on both 
$\sigsip$ (contribution proportional to $\sigsdp$ is negligible) and $\sigmav$. This is understood by the fact that DM density in the former case has already reached equilibrium while DM density in the latter case has not. Hence the search for neutrino signature from the Earth core can probe both cross sections.         

Model-independent sensitivity studies on IceCube detection of  DM induced neutrino signature from the Earth core  were reported in \cite{Delaunay:2008pc, Albuquerque:2011ma} for DM mass around TeV. In this work, we consider an extended DM mass range from few tens of GeV to  TeV. In the low mass range, our results can be compared with direct detection results from DAMA, CoGeNT, CRESST-II and XENON100 mentioned above. In the high mass range around TeV, our results can be compared with those from cosmic ray observations by PAMELA, Fermi-LAT and AMS02.
We study both muon track events and cascade events induced by neutrinos. We consider annihilation channels $\chi\chi\to \nu\bar{\nu}, \  
\chi\chi \to \tau^+\tau^-, \ W^+W^-, \ b\bar{b}$ for signature neutrino productions. For non-monochromatic modes, we note that $\chi\chi\to \tau^+\tau^-$ produces hardest neutrino spectrum while 
$\chi\chi \to b\bar{b}$ produces the softest one. We also include  $\nu_{\mu}\to \nu_{\tau}$ oscillations for  lower energy neutrinos.   

This paper is organized as follows. In Sec. II, we discuss the neutrino flux produced in the Earth core by DM annihilations. The procedure for calculating such a flux is outlined. In Sec. III, we discuss 
the track and shower event rates resulting from DM annihilations in the Earth core. The background event rates from atmospheric neutrino flux are also calculated. We adopt the effective areas published by IceCube observatory for event rate calculations. In Section IV, we present IceCube/DeepCore 5-year sensitivities for detecting DM induced neutrino signature from the Earth core. 
%We present such sensitivities in three different ways. 
We first fix the DM mass at two representative values, $m_{\chi}=50$ GeV and $m_{\chi}=2$ TeV. The IceCube/DeepCore  $2\sigma$ sensitivity for 5-year data taking is then presented as a curve on $(\sigmav, \sigsip)$ plane. Next, we fix the annihilation cross section $\sigmav$ at conservative values, $3\times 10^{-26}\rm{cm}^3 s^{-1}$ and $3\times 10^{-27}\rm{cm}^3 s^{-1}$. We then present IceCube/DeepCore sensitivities to spin-independent cross section $\sigsip$ as a function of  $m_{\chi}$ for different assumptions on dominant DM annihilation channels. Such sensitivities are then compared with existing  constraints  from direct detection experiments and that obtained from the IceCube/DeepCore search of DM annihilations in the Sun. Finally, we take different experimental bounds on $\sigsip$ as inputs to obtain different IceCube/DeepCore sensitivities to DM annihilation cross section on the $(m_{\chi}, \sigsip)$ plane for different annihilation channels. There are thus three scenarios for the input $\sigsip$:
(A) $\sigsip$ favored by DAMA and CRESST-II; (B) $\sigsip$ bound set by XENON100; (C) $\sigsip$  bound set by XENON1T (2017) \cite{Aprile:2012zx} assuming non-detection.  
We particularly compare  IceCube sensitivity to $\langle\sigma(\chi\chi\to \tau^+\tau^-)\upsilon\rangle$ to the favored range on the same quantity  implied by        
PAMELA, Fermi-LAT and AMS02. We note that neither $\chi\chi \to W^+W^-$ nor $\chi\chi \to b\bar{b}$ can simultaneously fit well to PAMELA and AMS02 \cite{Yuan:2013eja} if either channel is assumed to be dominant. We conclude in Sec. V.

\section{Neutrino flux from DM annihilation in the Earth core}
To facilitate our discussions, let us define
$dN^f_{\nu_i}/dE_\nu $ as the energy spectrum of $\nu_i$ produced  per DM annihilation $\chi\chi\to f\bar{f}$ in the Earth core. 
The differential DM neutrino flux of flavor $i$ on the Earth surface is then given by 
\begin{equation}
\frac{d\Phi^{\textrm{DM}}_{\nu_i}}{dE_\nu} =P_{\nu_j\to \nu_i}(E_\nu, D)
\frac{\Gamma_A}{4\pi D^2} \sum_{f}
B^{f}_{\chi}\frac{dN^f_{\nu_i}}{dE_\nu}
\label{eq:dNdE},
\end{equation}
where $\Gamma_A$ is the DM annihilation rate, 
$B^{f}_{\chi}$ is the branching ratio for the DM annihilation channel $\chi\chi\to f\bar{f}$, $D$ is the distance between the source and the detector,
and $P_{\nu_j\to \nu_i}(E_\nu, D)$ is the neutrino oscillation probability from the source to the detector.

To calculate $d\Phi^{\textrm{DM}}_{\nu_i}/dE_\nu$, 
we employ \texttt{WIMPSIM} \cite{Blennow:2007tw} with 
a total of 50000 Monte Carlo generated events.  
Although we are particularly interested in IceCube/DeepCore measurements, 
the DM neutrino flux from the Earth core is the same for all 
detector locations near the Earth surface due to 
the spherical symmetry.
The oscillation probability $P_{\nu_j\to \nu_i}(E_\nu, D)$ is calculated with the best fit neutrino oscillation parameters 
summarized in Table I of Ref.~\cite{Fogli:2012ua}, $\theta_{12}=33.65^{\circ}$, 
$\theta_{13}=8.93^{\circ}$, $\theta_{23}=38.41^{\circ}$, $\delta=1.08\pi$, 
$\delta m^2_{21}=7.54\times 10^{-5}$ $\rm{eV}^2$, and 
$\delta m^2_{31}=2.47\times 10^{-3}$ $\rm{eV}^2$. 

The DM annihilation rate $\Gamma_A$ can be determined by the following argument.
When the Earth sweeps through DM halo, the WIMP could 
collide with matter inside the Earth and lose its speed. 
If the WIMP speed becomes less than its escape velocity, the WIMP can be captured 
by Earth's gravitational force and then sinks into the core of Earth. 
After a long peroid of accumulation, WIMPs inside the core of Earth can begin to 
annihilate into Standard Model particles at an appreciable rate. Among the annihilation final states, neutrino can 
be detected by neutrino telescopes. 
Let $N(t)$ be the number of WIMPs in the Earth core at time $t$, we have
\begin{equation}
\frac{dN}{dt} = C_c - 2\Gamma_A (t)  -C_E N\label{eq:x2}, 
\end{equation}
where $C_c$ is the capture rate and $C_E$ is the evaporation rate. It has been shown that  WIMPS with masses between $5-10$ GeV may evaporate from the Earth~\cite{Krauss:1985aaa,Griest:1986yu,Gould:1987ir,Nauenberg:1986em}.
Since we are interested in the mass range $m_{\chi}> 10$ GeV, we neglect  $C_E$ in our discussions.  
The capture rate $C_c$ depends on DM-nuclei elastic scattering cross section which contains 
spin-dependent component $\sigsdp$ and spin-independent component $\sigsip$.   
The DM annihilation rate $\Gamma_A (t)$ is proportional to 
$N^2(t)$. One writes
\begin{equation}
\Gamma_A (t) = \frac{1}{2} \ C_A \, N^2 (t) \, . \label{eq:x1}  
\end{equation}
Taking into account the quasi-thermal distribution of WIMPs in the Earth core, 
the annihilation coefficient $C_A$ can be written as \cite{Berezinsky:1996ga}
\begin{equation}
C_A =\frac{\langle \sigma_a \, v \rangle}{V_{0}} \, \left( \frac{m_{\chi}}{20\,  \textrm{GeV}} \right)^{2/3}  , \label{eq:ca}
\end{equation}
where $V_{0}=2.3\times 10^{25}$ $\rm{cm}^3$ for the Earth.

By solving $N(t)$ in Eq.~(\ref{eq:x2}), we obtain \cite{Griest:1986yu} 
\begin{equation}
\Gamma_A(t)={C_c\over 2} \tanh^2\left({t\over \tau_A}\right),\label{eq:tanh}
\end{equation}
where $t$ is the age of the macroscopic body, for example $t = 4.5$ Gyr for Sun and Earth, while 
$\tau_{A}$ is the equilibration time scale, $\tau_{A}=(C_cC_A)^{-1/2}$.
Numerically  $\tanh^2(t/\tau_A)\to 1$ for ${t\over \tau_A}>2$. In such a case $\Gamma_A(t)=C_c/2$
so that the DM annihilation rate in the Earth core depends only on the capture rate $C_c$ and is independent of the annihilation cross section $\sigmav$. 

Since the heavy nuclei such as iron is abundant in the Earth core, 
the capture cross-section is enhanced due to its quadratic dependence on the nuclei atomic number. 
Therefore, the corresponding capture rate is given by \cite{Jungman:1995df}
\begin{equation}
C_c\propto \frac{\rho_{0}}{\rm{GeV}\,cm^{-3}} \times
\frac{\rm km\,s^{-1}}{\bar{v}}\times
\frac{\rm{GeV}}{m_{\chi}}\times 
\frac{\sigsip}{\rm{pb}}\times \sum_A F_A^{*}(m_\chi),
\end{equation}
with $\bar{v}$ the DM velocity dispersion, 
$\rho_0$ the local DM density, 
and $A$ the atomic number of chemical element in the Earth core.
$F_A^{*}(m_\chi)$ is the product of various factors 
including the mass fraction of element $A$, 
the gravitational potential for element $A$, 
kinematic suppression factor, form factor, and a factor 
of reduced mass. 
The explicit form of $F_A^{*}(m_\chi)$ is not 
essential for our discussions.  
It can be found, for instance, in Ref. \cite{Jungman:1995df}.
However, we like to stress that there are significant astrophysical uncertainties 
on the DM local density and its velocity distributions involved in $C_c$ \cite{Cirelli:2005gh}. 
In this work, we use approximate formulae given in \cite{Jungman:1995df}, 
which are adopted by \texttt{DarkSUSY}~\cite{Gondolo:2004sc}.

\section{DM signal and atmospheric background events}\label{sec:events}
Neutrino telescope such as IceCube detects neutrinos by measuring muon track and cascade events, which are  induced by neutrino-nucleon charged-current (CC) and 
neutral-current (NC) scatterings.  
We calculate neutrino event rate according to IceCube published  
neutrino effective area $A^{i,k}_{\textrm{eff}}(E_{\nu})$ \cite{Collaboration:2011ym} of the full IceCube 86-string detector,
where $k$ is interaction type for different neutrino flavor $i$. 
For example, for muon track events, $i$ is (anti-) muon neutrino and $k$ is (anti-) muon neutrino CC interaction.   
On the other hand, for cascade events, $i$ includes all three neutrino flavors. For (anti-) electron neutrinos and (anti-) tau neutrinos, $k$ runs over CC and NC interactions while
$k$ is exclusively NC interaction for  (anti-) muon neutrinos. 
The effective area accounts for the detection efficiency including the
neutrino-nucleon interaction probability, the energy loss of muon from its 
production point to the detector, and the detector trigger, 
and analysis efficiency.
Hence, the neutrino event rate from the Earth DM is given by
%%%%%%%%%
\begin{equation}
N_{\textrm{signal}}=\int_{E^{\rm th}}^{m_{\chi}}\sum_{i,k}\frac{d\Phi^{\textrm{DM}}_{\nu_i}}{dE_\nu}A^{i,k}_{\textrm{eff}}(E_{\nu})dE_{\nu}d\Omega\,,
\label{eq:Nevents2}
\end{equation}
%%%%%%%%%
where $d \Phi^{\textrm{DM}}_{\nu_i}/dE_{\nu}$ is the differential neutrino flux in the vicinity of detector for a given neutrino flavor $i$, which is given by Eq.~(\ref{eq:dNdE}), and the index $k$ can be either NC or CC interaction. The effective area can be defined as
$A^{i,k}_{\textrm{eff}}(E_{\nu})\approx\rho_{\textrm{ice}}N_{A}\sigma^{k}_{\nu_{i} N}(E_{\nu})V^{i,k}_{\textrm{eff}}(E_{\nu})$ \cite{Buckley:2009kw}, where
$ \rho_{\textrm{ice}}=0.9 $  $\rm{g}/\rm{cm}^3$ is the density of ice, $ N_{A}=6.022\times 10^{23}$ $ \rm{g}^{-1} $ is the Avogadro number, $\sigma^{k}_{\nu_{i} N}(E_{\nu})$ is the neutrino-nucleon
cross-section, and $ V^{i,k}_{\textrm{eff}}(E_{\nu}) $ is the effective volume of IceCube for different neutrino-nucleon interaction events.
In this work, we shall take the detector threshold energy $E^{\rm th}$ for IceCube/DeepCore 
as 10 GeV and 100 GeV, respectively. 
To compute the rate for $\nu_{\mu}$ track events, we shall use the effective area (DeepCore+IceCube Trigger) given by Ref.~\cite{Collaboration:2011ym}.

Unlike the calculation of  track event rate, which requires only one effective area, the calculation of cascade event rate requires 5 different effective areas. 
The cascade-event effective area given by Ref.~\cite{Collaboration:2011ym} is only for $\nu_e$ CC interaction events. Here we also adopt the effective area marked as DeepCore+IceCube 
Trigger for estimating the $\nu_e$ event rate.  
To obtain effective areas for other cascade events,  we perform the rescaling 
\begin{equation}
A^{i,k}_{\textrm{eff}}(E_{\nu})= 
A^{\nu_{e},\rm{CC}}_{\textrm{eff}}(E_{\nu})\times\frac{\sigma^{k}_{\nu_{i} N}(E_{\nu})}{\sigma^{\rm{CC}}_{\nu_e N}(E_{\nu})} 
\times \frac{ V^{\nu_{e},\rm{CC}}_{\textrm{eff}}(\langle y\rangle\cdot E_{\nu}) } {V^{\nu_{e},\rm{CC}}_{\textrm{eff}}(E_{\nu})}, 
\end{equation}
where $\langle y\rangle$ is the averaged fraction of 
neutrino energy $E_\nu$ converted into shower energy after a neutrino-nucleon CC or NC interaction. 
One has $\langle y\rangle=0.3$ for NC interactions of $\nu_e$ and $\nu_{\mu}$ while  
$\langle y\rangle=1$ for $\nu_e$ CC interaction~\cite{Beacom:2004}. 
For $\nu_{\tau}$, the fraction $\langle y\rangle$ resulting from
$\nu_\tau$ CC interaction and subsequent tau-lepton decay is approximately $0.6\times\langle y_h\rangle + 0.4$ 
where $\langle y_h\rangle$  
is the energy fraction of $\nu_\tau$ taken by hadrons in $\nu_\tau$-nucleon CC interaction~\cite{Gandhi:1998}. 
The factor 0.4 is the visible energy fraction in tau lepton decays, 
which can be estimated by using PYTHIA \cite{Sjostrand:2006za}. It should be noted that the final effective area/effective volume for cascade events may be significantly reduced due 
to the background rejection cut. In a recent IceCube 79-string result on atmospheric $\nu_e$ flux measurement~\cite{Aartsen:2013}, the final effective volume for $\nu_e$ events is much smaller than the one at the DeepCore filter level for lower energies while such a difference is within an order of magnitude for $E_{\nu} \gtrsim 10^3$ GeV, as can be seen from Fig.~2 of that paper. Since our sensitivity calculations for cascade events are based upon effective areas in Ref.~\cite{Collaboration:2011ym}, we only present  IceCube/DeepCore sensitivities for $E^{\rm th}=100$ GeV in the case of cascade events.

For calculating the atmospheric  (ATM) background event rates, 
we use the atmospheric neutrino flux  
taken from Ref.~\cite{Honda:2006qj}. 
Because the intrinsic ATM $\nu_{\tau}$ flux is negligible, 
we only consider ATM $\nu_e$ and $\nu_{\mu}$ fluxes at the production point.  
It is essential to include the effect of neutrino transmissions through the Earth. 
%We therefore introduce neutrino transmission probability through the Earth, $P_{\textrm{earth}}(E_{\nu}, \theta_{\nu} )$, where $ \theta_{\nu} $ is the neutrino angle with respect to the detector coordinate system. 
Since the Earth becomes opaque to neutrinos only for $E_{\nu}>40$ TeV~\cite{Gandhi:1998}, the neutrino transmission in our interested energy range is essentially the neutrino oscillation effect.
Hence the ATM background event rate is given by   
%%%%%%%%%
\begin{equation}
N_{\textrm{background}}=\int_{E^{\rm th}}^{E_{\rm max}}\sum_{i,k}\frac{d\Phi^{\textrm{ATM}}_{\nu_i}}{dE_{\nu}}A^{i,k}_{\textrm{eff}}(E_{\nu})dE_{\nu}d\Omega\, ,
\label{eq:Neventb1}
\end{equation}
%%%%%%%%%
where $d\Phi^{\textrm{ATM}}_{\nu_i}/dE_{\nu}$ is ATM neutrino flux in the vicinity of the detector. Such a flux is given by
%%%%%%%%%
\begin{equation}
\frac{d\Phi^{\textrm{ATM}}_{\nu_i}}{dE_{\nu}}=\frac{d \Phi^{\textrm{ATM}}_{0,\nu_j}}{dE_{\nu}}P_{\nu_{j}\rightarrow\nu_{i}}\left( E_{\nu},L(\theta_{\nu})\right) \,,
\label{eq:Neventb2}
\end{equation}
%%%%%%%%%
where $d \Phi^{\textrm{ATM}}_{0,\nu_j}/dE_{\nu}$ is the ATM neutrino flux at the source, $ P_{\nu_{j}\rightarrow\nu_{i}}(E_{\nu},L(\theta_{\nu})) $ 
is the neutrino oscillation probability  with
$ L(\theta_{\nu})$ the neutrino traversing distance 
through the Earth along the direction of $ \theta_{\nu}$. 
For comparing with the signal event rate induced by Earth DM annihilation with DM mass $m_{\chi}$, $E_{\rm max}$ is taken as $m_{\chi}$ in Eq.~(\ref{eq:Neventb1}). 
We remark that $E_{\rm max}$ should differ from $m_\chi$ in practice due to the energy resolution effect. However the impact of energy resolution on our sensitivity estimation will be shown insignificant
in the next section.

%%%%%%%%%%%%%%%%%%%%%%%%%   F   I   G   U   R   E   %%%%%%%%%%%%%%%%%%%%%%%%%%%%
\begin{figure}[ht!]
\centering
\includegraphics[width=0.49\textwidth]{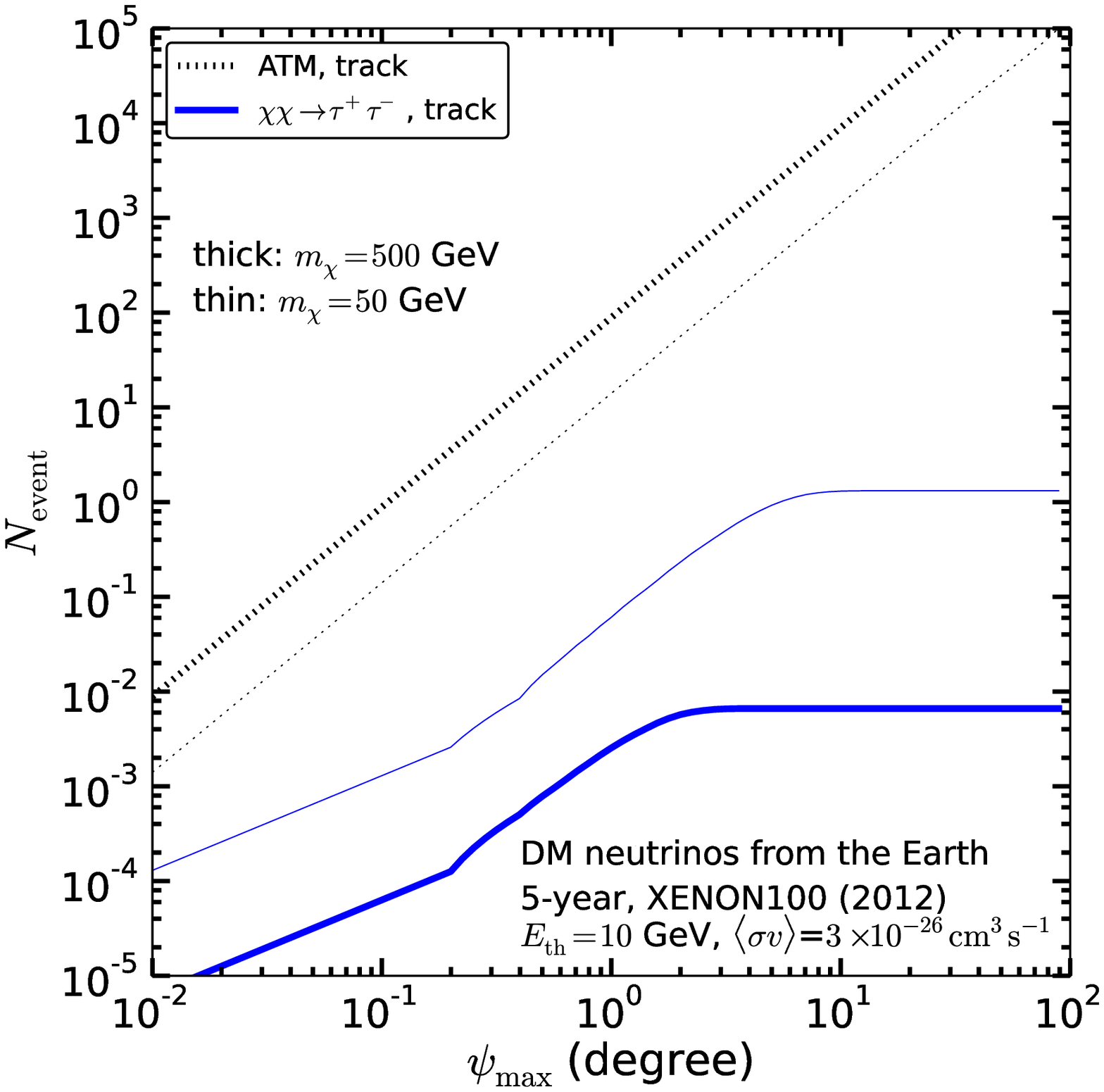}
\includegraphics[width=0.49\textwidth]{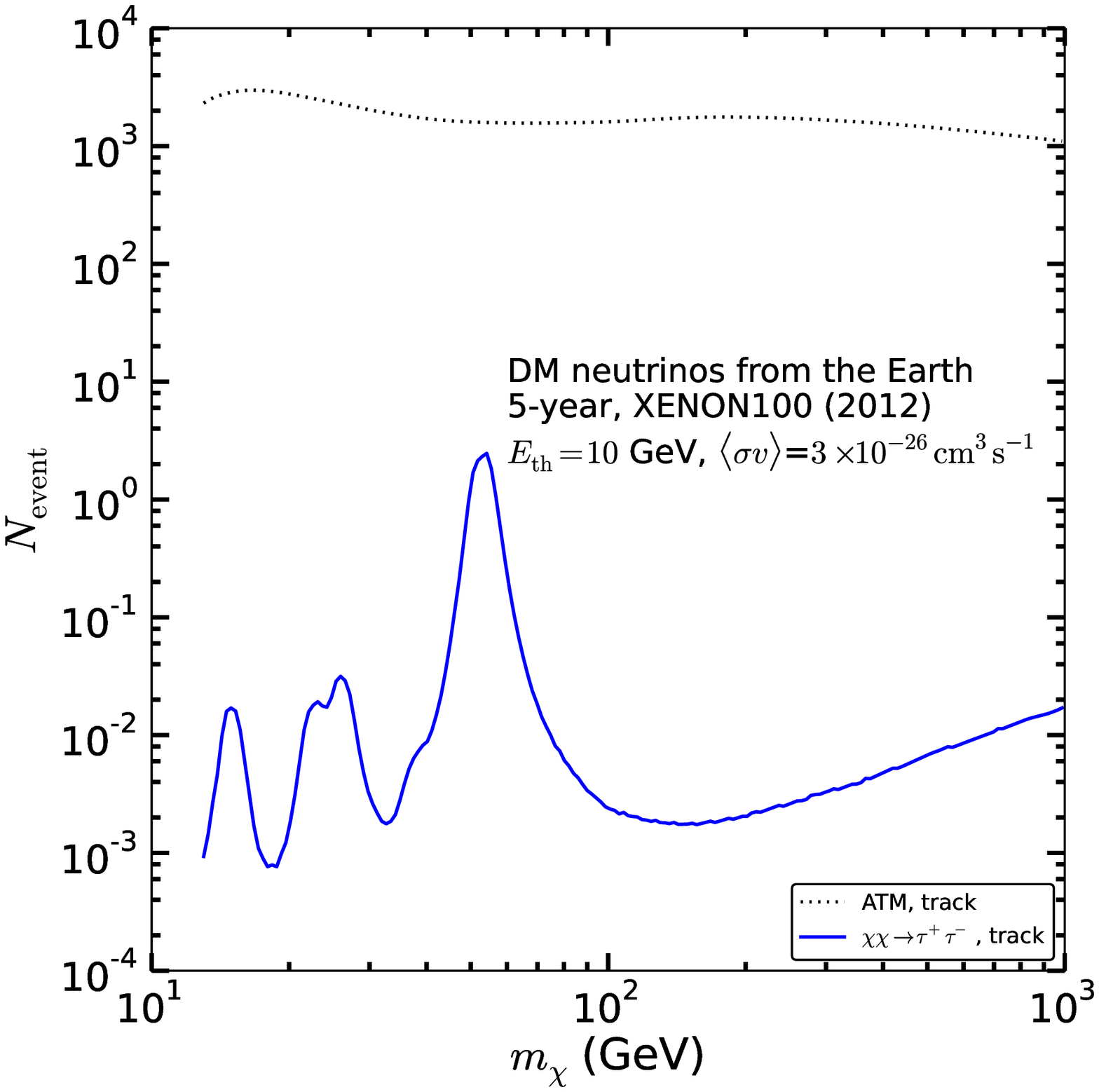}
\caption[]{The events collected in 5 year. We display the total events 
in the detector as function of $\psi_{\rm{max}}$ (left) and $m_{\chi}$ (right). 
}
\label{fig:Nevents}
\end{figure} 
%%%%%%%%%%%%%%%%%%%%%%%%%%%%%%%%%%%%%%%%%%%%%%%%%%%%%%%%%%%%%%%%%%%%%%%%%%%%%%%%

In Fig. \ref{fig:Nevents}, we show the number of neutrino events in 5 years from DM annihilations ($\chi\chi\to \tau^+\tau^-$) and ATM background as  functions of the
maximum open angle $\psi_{\rm{max}}$ (left) and the DM mass $m_{\chi}$ (right), respectively. 
Here, we take $\sigmav=3\times 10^{-26}\rm{cm}^3 s^{-1}$ and $\sigsip$ the 90\% CL upper limit 
from XENON100.  
From Eq.~(\ref{eq:ca}), we can see the annihilation coefficient $C_A$ is  inverse proportional 
to a $m_{\chi}$ dependent effective volume 
$V_{\rm{eff}}=V_{0}\left( 20\, \textrm{GeV}/m_{\chi} \right)^{2/3}$, 
which describes the volume of dark matter occupation in the Earth core.
Hence it is seen from the left panel of Fig. \ref{fig:Nevents} that  $N_{\rm{events}}$ reaches to the maximum for sufficiently  
large $\psi_{\rm{max}}$ that can cover the entire DM populated region in the Earth core. 
The critical value of $\psi_{\rm{max}}$ for covering the DM populated region in the Earth core  
is a function of $m_{\chi}$, which we denote as $\psi^{\rm{c}}_{\rm{max}}(m_\chi)$. We have 
%%%%%%%%%
\begin{equation}
\psi^{\rm{c}}_{\rm{max}}(m_\chi)={\rm max}\left[\sin^{-1}\left(\frac{1}{R_\oplus}\times 
\left(\frac{3 V_{\rm{eff}}(m_\chi)}{4\pi}\right)^{\frac{1}{3}}\right),1^{\circ}\right], 
\label{eq:phimax}
\end{equation}
%%%%%%%%%
where $R_\oplus$ the radius of the Earth. The $1^{\circ}$ on the right hand side of the equation is to ensure a minimal open angle of $1^{\circ}$.
We have seen that $V_{\textrm{eff}}(m_{\chi})$ decreases as $m_{\chi}$ increases. 
Hence,  $\psi^{\rm{c}}_{\rm{max}}$ for $m_{\chi}=500$ GeV is smaller than $\psi^{\rm{c}}_{\rm{max}}$ for $m_{\chi}=50$ GeV, 
as can be seen from the left panel of Fig.~\ref{fig:Nevents}.
The right panel of Fig.~\ref{fig:Nevents} shows DM and ATM background 
event numbers as functions of $m_{\chi}$ where  
$\psi_{\rm{max}}$ for each $m_{\chi}$ is taken to be  $\psi^{\rm{c}}_{\rm{max}}(m_{\chi})$.
The $N_{\rm events}$ for DM signal peaks at three different values of $m_{\chi}$. This is due to the enhancement of capture rate when $m_{\chi}$ is close 
to the mass of any  dominantly populated nuclei  in the Earth core. In fact, the three peaks from small to large $m_{\chi}$ correspond to the resonant capture by oxygen, Mg/Si, and Fe/Ni, respectively. 
Effects of these resonant capture peaks have been studied 
in inert doublet  \cite{Andreas:2009hj}
and supersymmetry neutralino DM models~\cite{Bottino:2004qi,Niro:2009mw}. 
In Ref.~\cite{Bruch:2009rp}, the authors also included the effect 
from dark disc and found that the search for 
DM annihilation in the Earth can have the same level of sensitivity 
as the search for DM annihilation in the Sun for $m_\chi\lesssim 100$ GeV.
  
%%%%%%%%%%%%%%%%%%%%%%%%%%%%%%%%%%%%%%%%%%%%%%%%%%%%%%%%%%%%%%%%%%%%%%%%%%%%%%%%
% Result Sec.
%%%%%%%%%%%%%%%%%%%%%%%%%%%%%%%%%%%%%%%%%%%%%%%%%%%%%%%%%%%%%%%%%%%%%%%%%%%%%%%%

\section{Result}\label{sec:result}
We present the sensitivity as a $2\sigma$ detection significance in 5 years, calculated with the
standard formula 
\begin{equation}
\frac{\rm{DM\,\, signal}}{\sqrt{\rm{ATM\,\, background}}}=2.0. 
\label{eq:sensitivity}
\end{equation}
The ATM here is the number of atmospheric background events, 
which we calculate with the flux data from Ref.~\cite{Honda:2006qj}. 
The right hand side of Eq.~(\ref{eq:sensitivity}) 
refers to the 2$\sigma$ detection significance.

As mentioned earlier, the effect of detector energy resolution on the sensitivity estimation needs to be understood. Here we assume IceCube can determine the neutrino energy 
with the energy resolution $\sigma_{E}/E=50\%$~\cite{Albuquerque:2011ma}.   
With this energy resolution, we may compare the ATM track event rate in  Eq.~(\ref{eq:Neventb1}) with $E_{\rm max}=m_{\chi}$ 
and the similar event rate with $E_{\rm max}=3m_{\chi}/2$, which has taken into account the $50\%$ energy resolution.    
We have found that, for $E^{\rm th}=10$ GeV and $12<E_{\rm max}/\rm{GeV}<10^3$, the ratio of the ATM background event rate with 
$E_{\rm max}=3m_{\chi}/2$ to that with $E_{\rm max}=m_{\chi}$ varies between 2.7 and unity. From Eq.~(\ref{eq:sensitivity}), we can see that 
the magnitude of 
DM annihilation cross section which the detector can probe is proportional to the square root of the ATM background event number. 
Thus the IceCube sensitivity to $\langle \sigma\upsilon\rangle$ with track events is changed only slightly by a factor $f$ in the range $1<f<1.65$. We expect similar effect for cascade events.
Therefore, the effect of energy resolution on our sensitivity estimation is insignificant.

%%%%%%%%%%%%%%%%%%%%%%%%%   F   I   G   U   R   E   %%%%%%%%%%%%%%%%%%%%%%%%%%%%
\begin{figure}[ht!]
\centering
\includegraphics[width=0.49\textwidth]{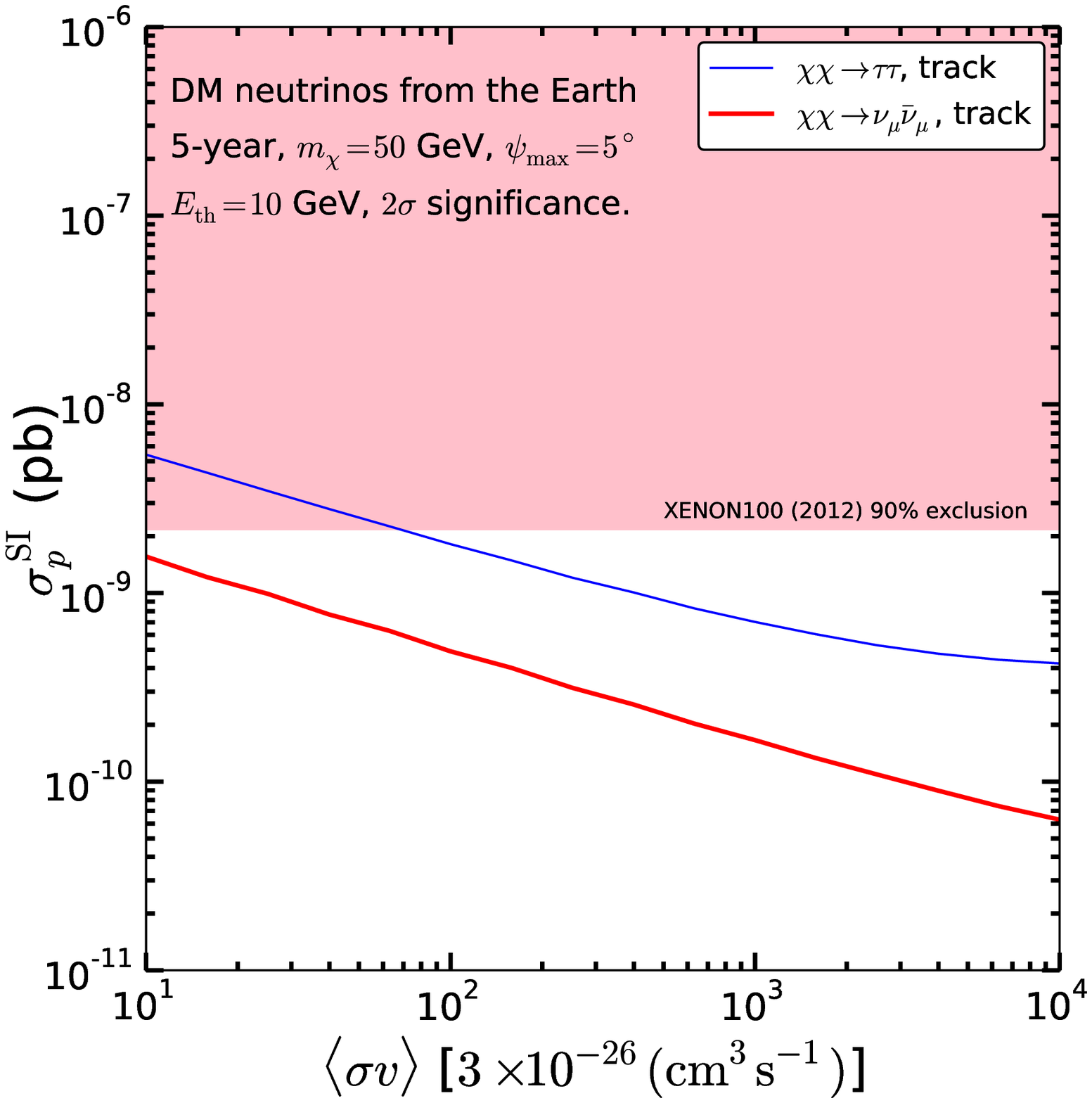}
\includegraphics[width=0.49\textwidth]{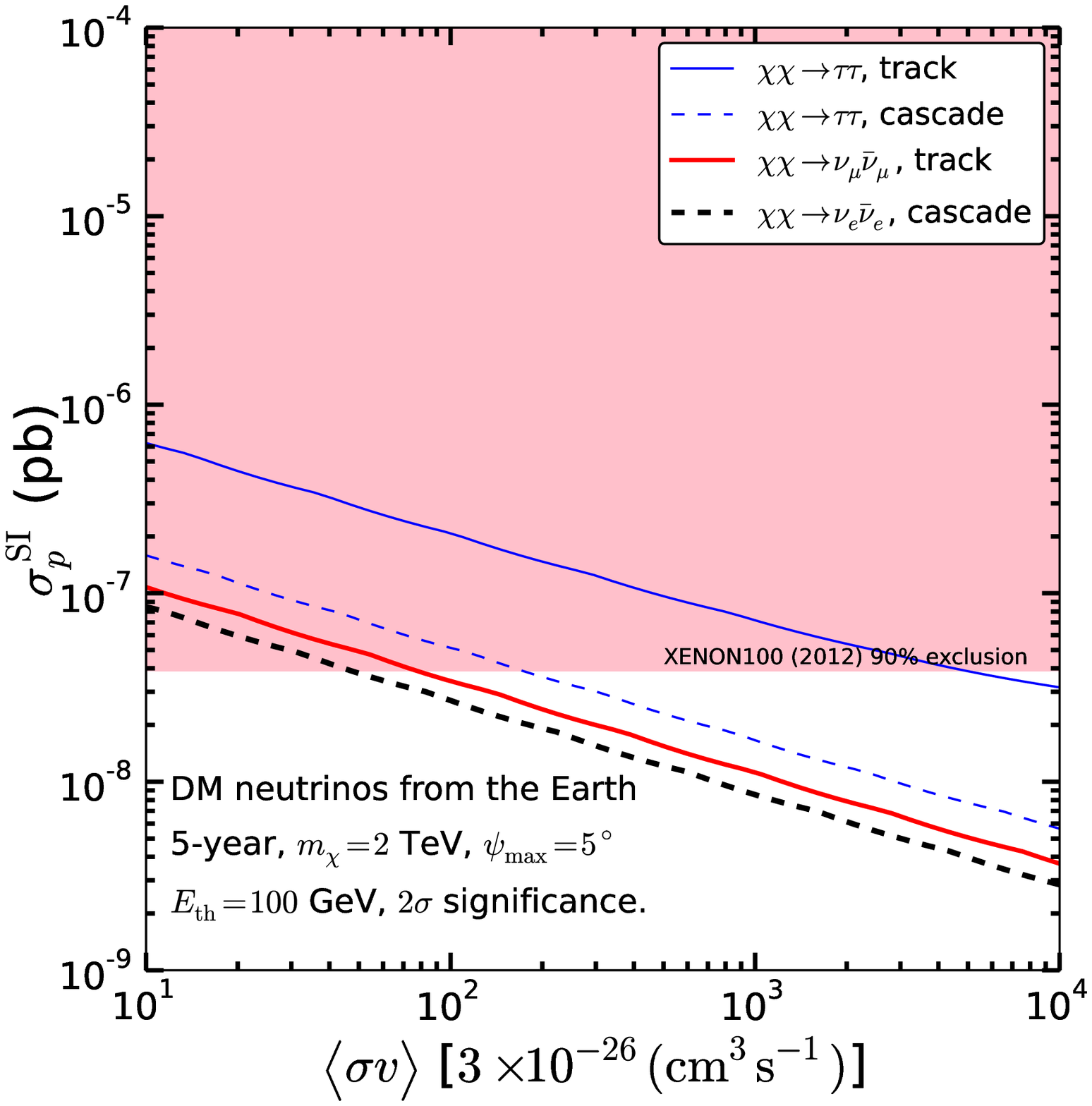}
\caption[]{The 5 year sensitivity in $2\sigma$ significance on 
($\sigmav$, $\sigsip$) plane. In the left panel, the DM mass is 50 GeV 
and the threshold energy is 10 GeV. 
The DM mass is 2 TeV  and threshold energy is taken to be 
100 GeV in the right panel. 
Solid lines are sensitivities with 
track events while dashed lines are sensitivities with cascade events. 
The pink shaded area represents the 90\% exclusion from XENON100 (2012) result.
}
\label{fig:sigsip_BF}
\end{figure} 
%%%%%%%%%%%%%%%%%%%%%%%%%%%%%%%%%%%%%%%%%%%%%%%%%%%%%%%%%%%%%%%%%%%%%%%%%%%%%%%%

In Fig. \ref{fig:sigsip_BF}, we show the 5 year sensitivity of the full IceCube 86-string detector to Earth DM signal on ($\sigmav$, $\sigsip$) plane. 
We consider annihilation channels $\chi\chi\to \nu_e\bar{\nu}_e, \ \nu_{\mu}\bar{\nu}_{\mu}, \ \textrm{and} \ \tau^+\tau^-$. The first 
annihilation mode produces cascade events while the second and third channels produce both track and cascade events due to 
$\nu_{\mu}\to \nu_{\tau}$ oscillations. Each annihilation channel is assumed to be dominant when IceCube/DeepCore  sensitivity to that channel is derived.     
The open angle $\psi_{\rm max}$ for collecting events from the Earth core is taken to be $5^{\circ}$ for both track and cascade events. This is 
reasonable for track events and achievable for high energy cascade events~\cite{Auer} 
such as those produced  with $m_{\chi}=2$ TeV described in the right panel. Since we only consider  cascade events with $E^{\rm th}=100~\textrm{GeV}$, the 
sensitivities with cascade events are only presented in the right panel. In the right panel,
we can see that $\chi\chi\to \nu_e\bar{\nu}_e$ channel is most sensitive to $\sigsip$. In particular, this channel provides better sensitivity than 
$\chi\chi\to \nu_{\mu}\bar{\nu}_{\mu}$ channel due to relatively less ATM background events.

We note that the 5 year sensitivity curve on ($\sigmav$, $\sigsip$) plane  is almost linear 
in logarithmic scale so that $\sigsip$ approximates to $\sigmav^{-k}$ with slope $-k$.       
As $\sigmav$ increases, a smaller scattering cross section $\sigsip$ is sufficient to achieve the same detection significance. 
However, Eq.~(\ref{eq:x1}) implies that this trend cannot continue indefinitely.
As $\tanh(\frac{t}{\tau_A})$ is driven to the plateau by a sufficiently large $\sigmav$, $C_c$ must approach to
a constant value for maintaining the same annihilation rate $\Gamma_A$. This then  implies that $\sigsip$ also approaches to a constant value. 
In the reverse direction where $\sigmav$ decreases, a larger $\sigsip$ is required to achieve the same detection significance. 
On the other hand, the XENON100 limit (pink shaded region) 
eventually sets the upper bound for $\sigsip$. This constraint is clearly seen for $\chi\chi\to \tau^+\tau^-$ channel with $m_\chi=2$ TeV. 
Due to XENON100 limit, 
such a channel cannot produce enough neutrino events in IceCube for reaching $2\sigma$ sensitivity in 5 years, unless 
the boost factor for $\sigmav$ is larger than $1000$.

It is clear that XENON100 limit sets a $m_\chi$-dependent bound on $\sigmav$ for each annihilation channel. 
However, if one takes different experimental bounds on $\sigsip$, the prospect of observing neutrinos from the Earth core differs drastically.  
There are thus three possibilities:

\noindent
Case A: Neutrino observation implied by DAMA and CRESST-II favored parameter space;\\  
Case B: $\sigmav$ exclusion limits implied by XENON100 bound on $\sigsip$;\\
Case C: Pessimistic scenario by assuming non-detection of XENON1T (2017).   

Below we shall discuss in turn these three cases.

%%%%%%%%%%%%%%%%%%%%%%%%%%%%%%%%%%%%%%%%%%%%%%%%%%%%%%%%%%%%%%%%%%%%%%%%%%%%%%%%
\subsection{Neutrino observation implied by DAMA and CRESST-II}

%%%%%%%%%%%%%%%%%%%%%%%%%   F   I   G   U   R   E   %%%%%%%%%%%%%%%%%%%%%%%%%%%%
\begin{figure}[p]
\centering
\includegraphics[width=0.49\textwidth]{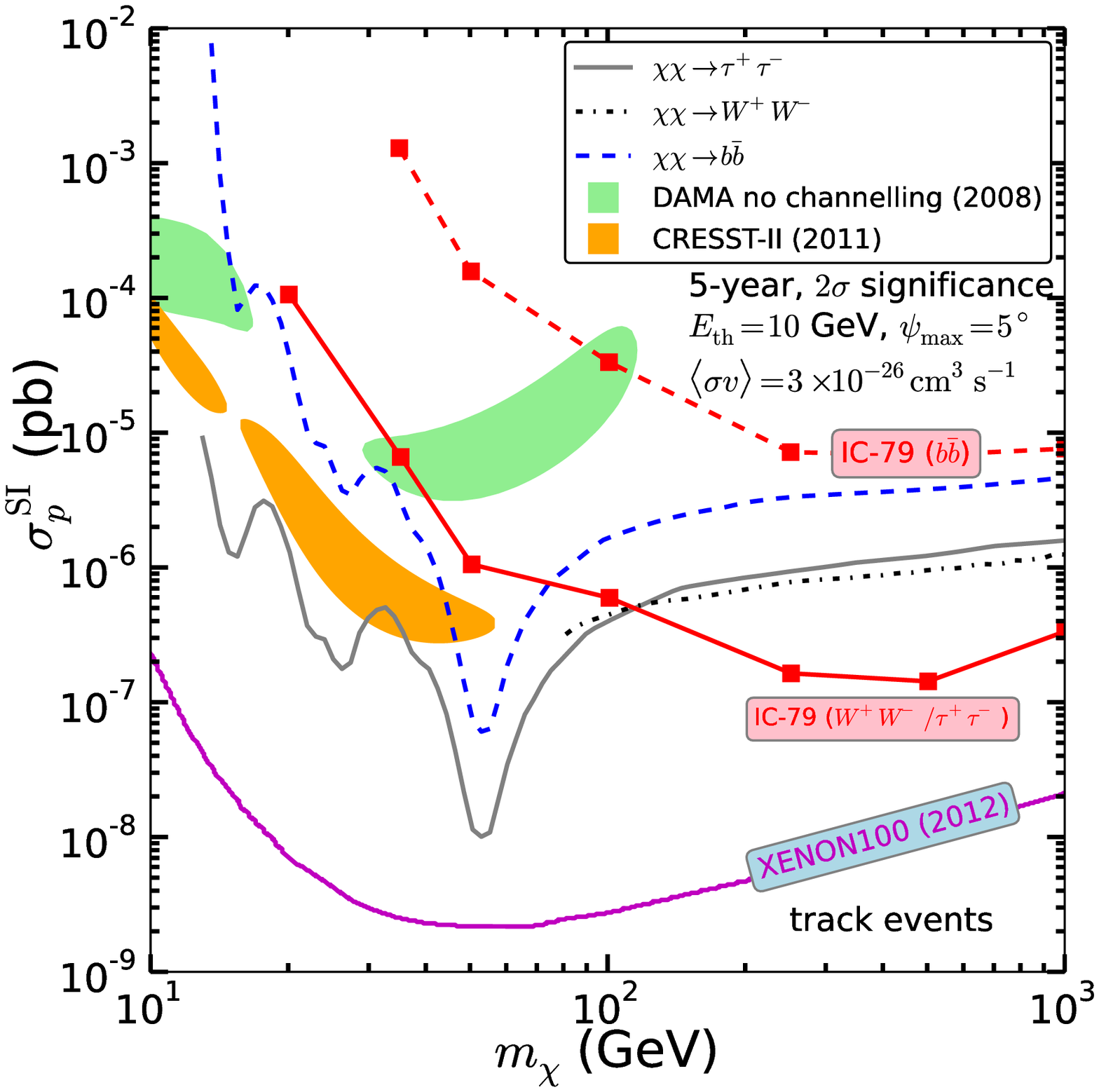}
\includegraphics[width=0.49\textwidth]{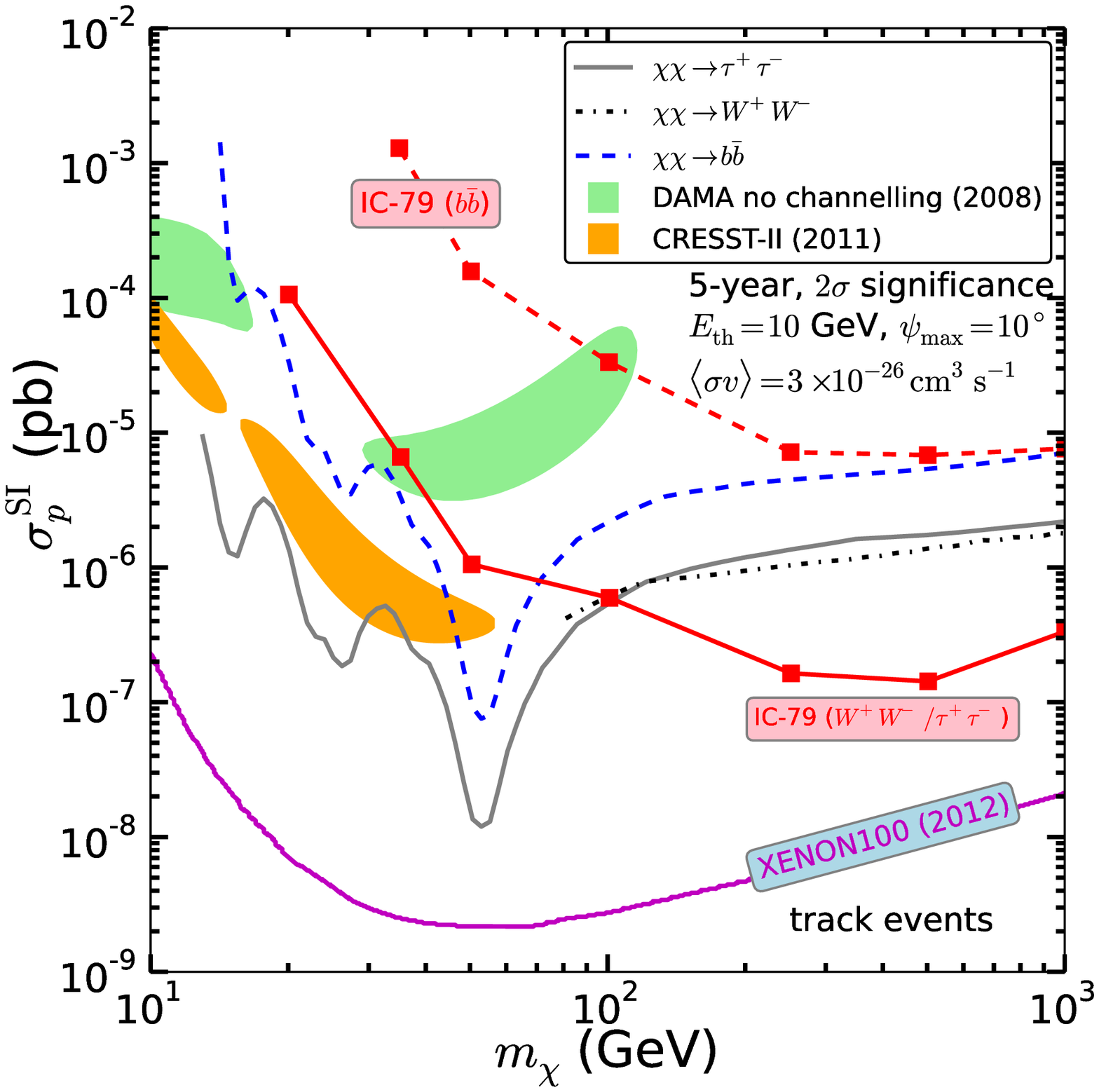}\\
\includegraphics[width=0.49\textwidth]{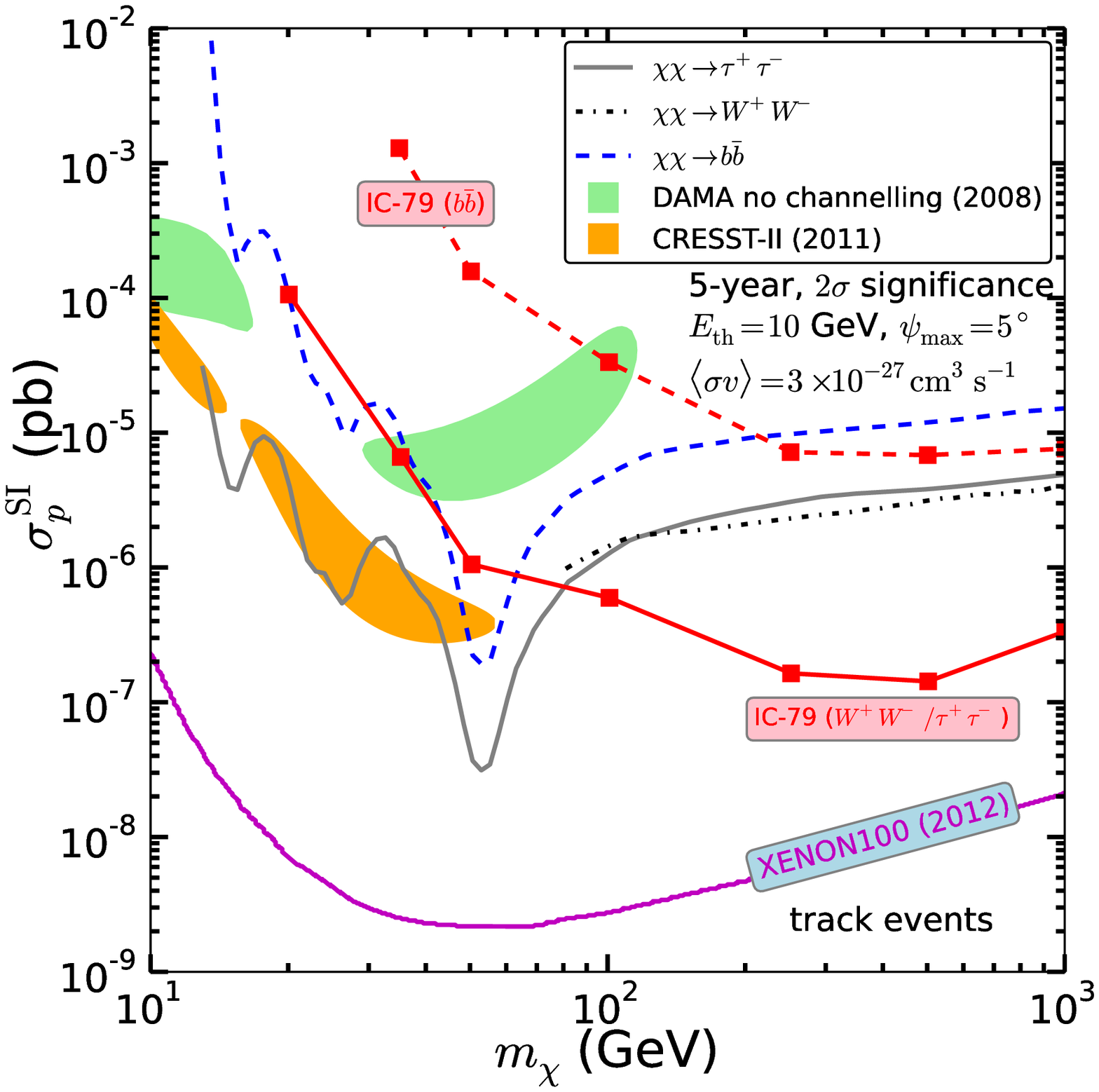}
\includegraphics[width=0.49\textwidth]{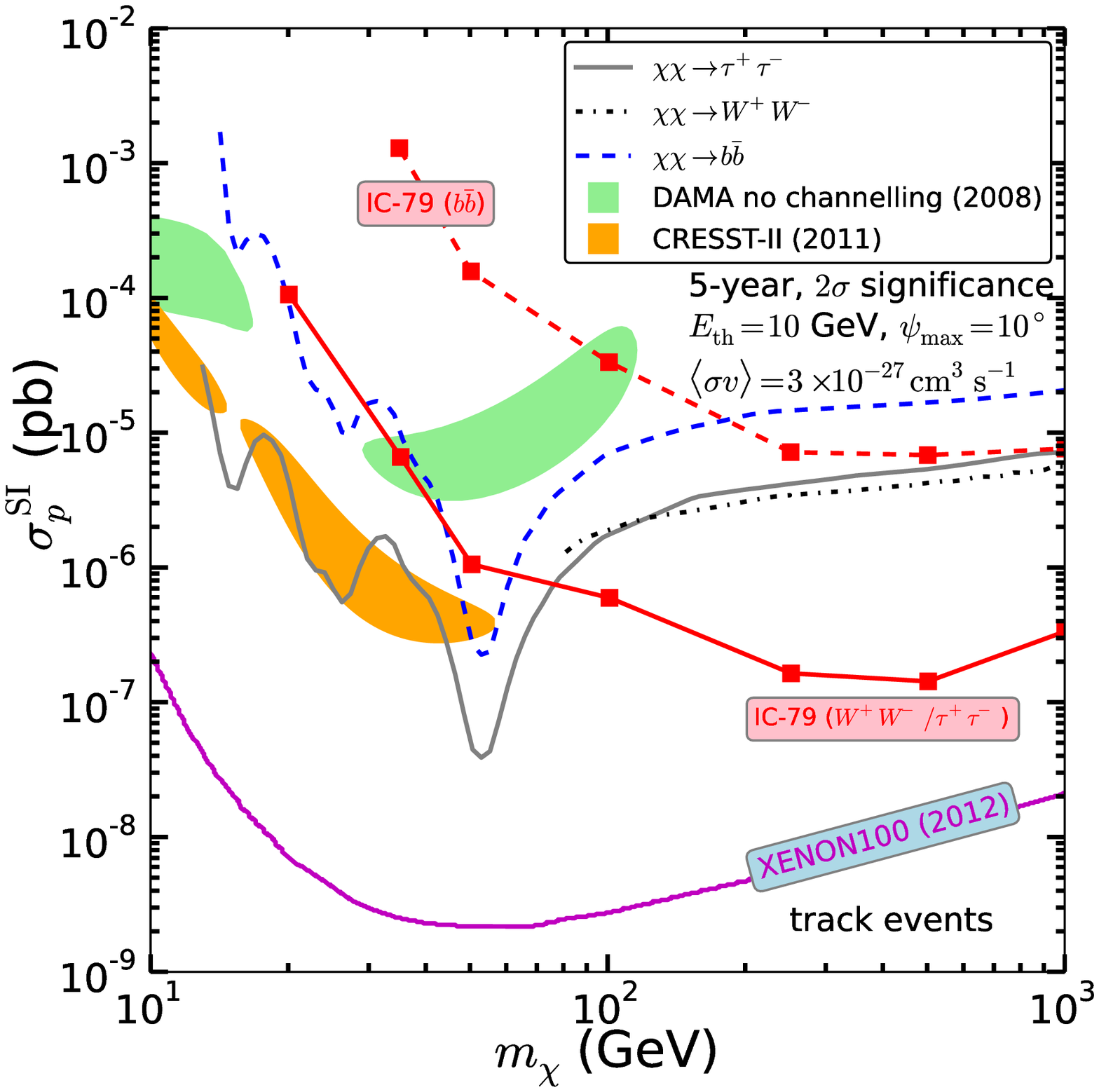}\\
\caption[]{The IceCube/DeepCore 5 year sensitivity curves on the  ($m_\chi$, $\sigsip$) plane   
for $\chi\chi\to \tau^+\tau^-, \ W^+W^-$, and $b\bar{b}$ annihilation channels.  
The upper figures are based on $\sigmav= 3\times 10^{-26} \rm{cm}^3 s^{-1}$ while 
the lower figures are based on $\sigmav= 3\times 10^{-27} \rm{cm}^3 s^{-1}$. 
The track-event sensitivities with $\psi_{\rm max}=5^{\circ}$ are presented in figures on the left column and 
that with $\psi_{\rm max}=10^{\circ}$ are presented in figures on the right column. The energy threshold is taken to be 10 GeV. 
The IceCube 79-string upper limits on $\chi\chi\to b\bar{b}$ 
(red dashed-squared line) and $\chi\chi \to W^+W^-/\tau^+\tau^-$ (red solid-squared line) 
from the search for DM-induced neutrino signature from the Sun are also shown for comparison~\cite{Aartsen:2012kia}.    
}
\label{fig:mx_sigsip}
\end{figure} 
%%%%%%%%%%%%%%%%%%%%%%%%%%%%%%%%%%%%%%%%%%%%%%%%%%%%%%%%%%%%%%%%%%%%%%%%%%%%%%%% 

In Fig.~\ref{fig:mx_sigsip}, we present IceCube/DeepCore sensitivities to 
$\chi\chi\to \tau^+\tau^-,  \ W^+W^-$, and $b\bar{b}$ annihilations in the Earth core on the 
($m_\chi$, $\sigsip$) plane for track events with different $\psi_{\rm{max}}$ and $\sigmav$ with $E^{\rm th}=10$ GeV.
The sensitivities to $\chi\chi\to W^+W^-$ (black dashed-doted) and 
$\chi\chi\to \tau^+\tau^-$ (grey solid) are comparable, albeit the $W^+W^-$ channel only opens at 
$m_\chi>m_W$. 
The experimental upper limits of $\chi\chi\to b\bar{b}$ (red dashed-squared line) 
and $\chi\chi\to W^+W^-/\tau^+
\tau^-$ (red solid-squared line) 
are taken from IceCube 79-string result on the search for muon neutrino events induced by DM annihilations in 
the Sun~\cite{Aartsen:2012kia}.  We note that the constraint on $\chi\chi\to W^+W^-/\tau^+\tau^-$  
stands for a constraint on $\chi\chi\to W^+W^-$ for $m_\chi> m_W$ and a constraint on $\chi\chi\to \tau^+\tau^-$
for  $m_\chi< m_W$.    
For $m_\chi\lesssim 100$ GeV, it is seen that $\sigsip$  
can be better probed by detecting DM induced neutrino signature from the Earth core than that from the Sun. 

The favoured region of DAMA at higher $m_\chi$ is not compatible 
with IceCube 79-string constraint on  $\chi\chi\to \tau^+\tau^-$. This region can also be probed by searching for neutrinos from  
$\chi\chi\to \tau^+\tau^-$ and $b\bar{b}$ annihilations in the Earth core.   
In fact, the search for muon track events induced by $\chi\chi\to \tau^+\tau^-$  with $\sigmav=3\times 10^{-26} \rm{cm}^3 s^{-1}$ 
can probe the full allowed region of DAMA and most of the allowed region of CRESST-II. 
With a 10 times smaller $\sigmav$,  the full allowed region of DAMA can still be probed by the same annihilation channel.

%%%%%%%%%%%%%%%%%%%%%%%%%   F   I   G   U   R   E   %%%%%%%%%%%%%%%%%%%%%%%%%%%%
\begin{figure}[ht!]
\centering
\includegraphics[width=0.49\textwidth]{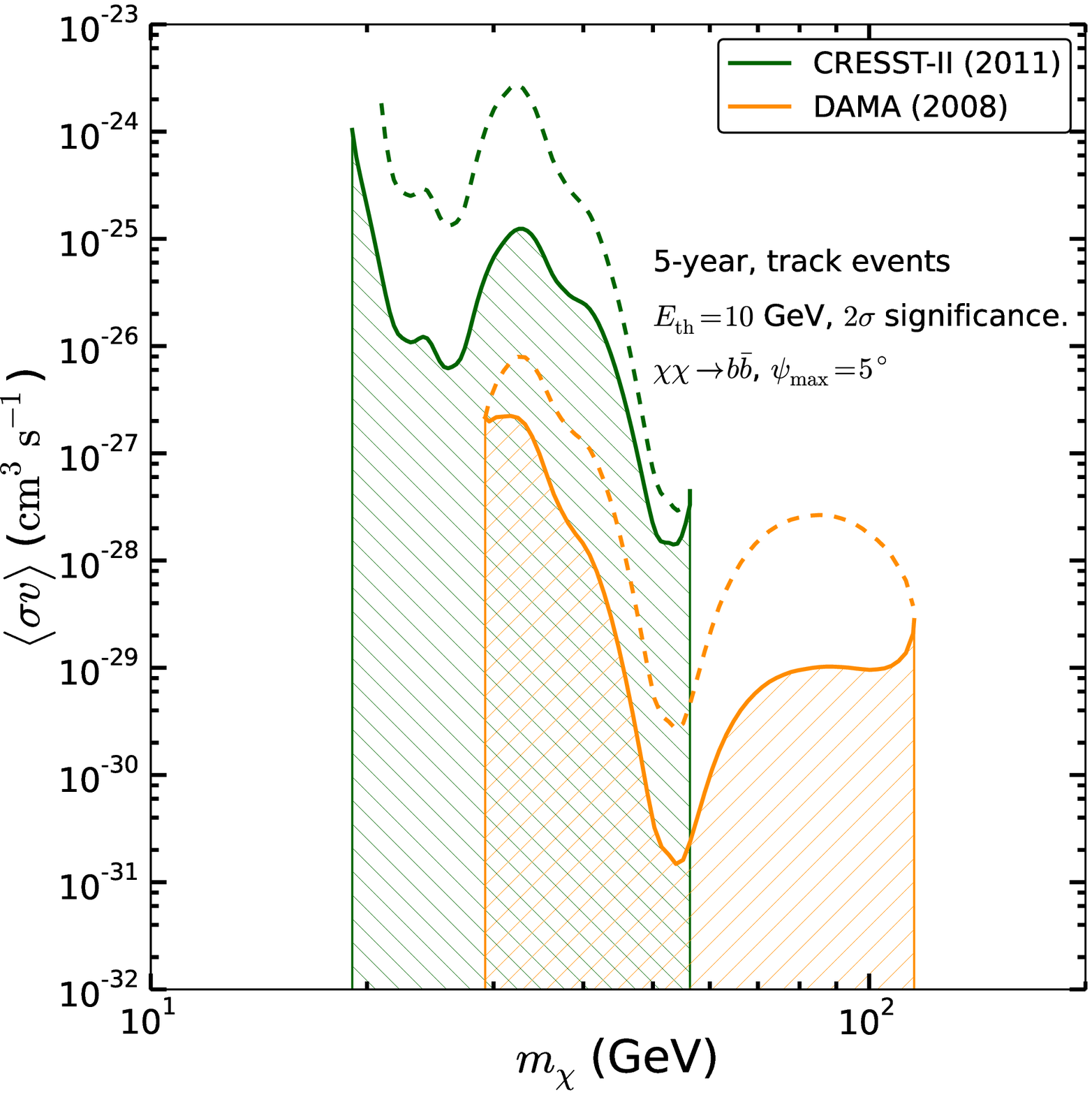}
\includegraphics[width=0.49\textwidth]{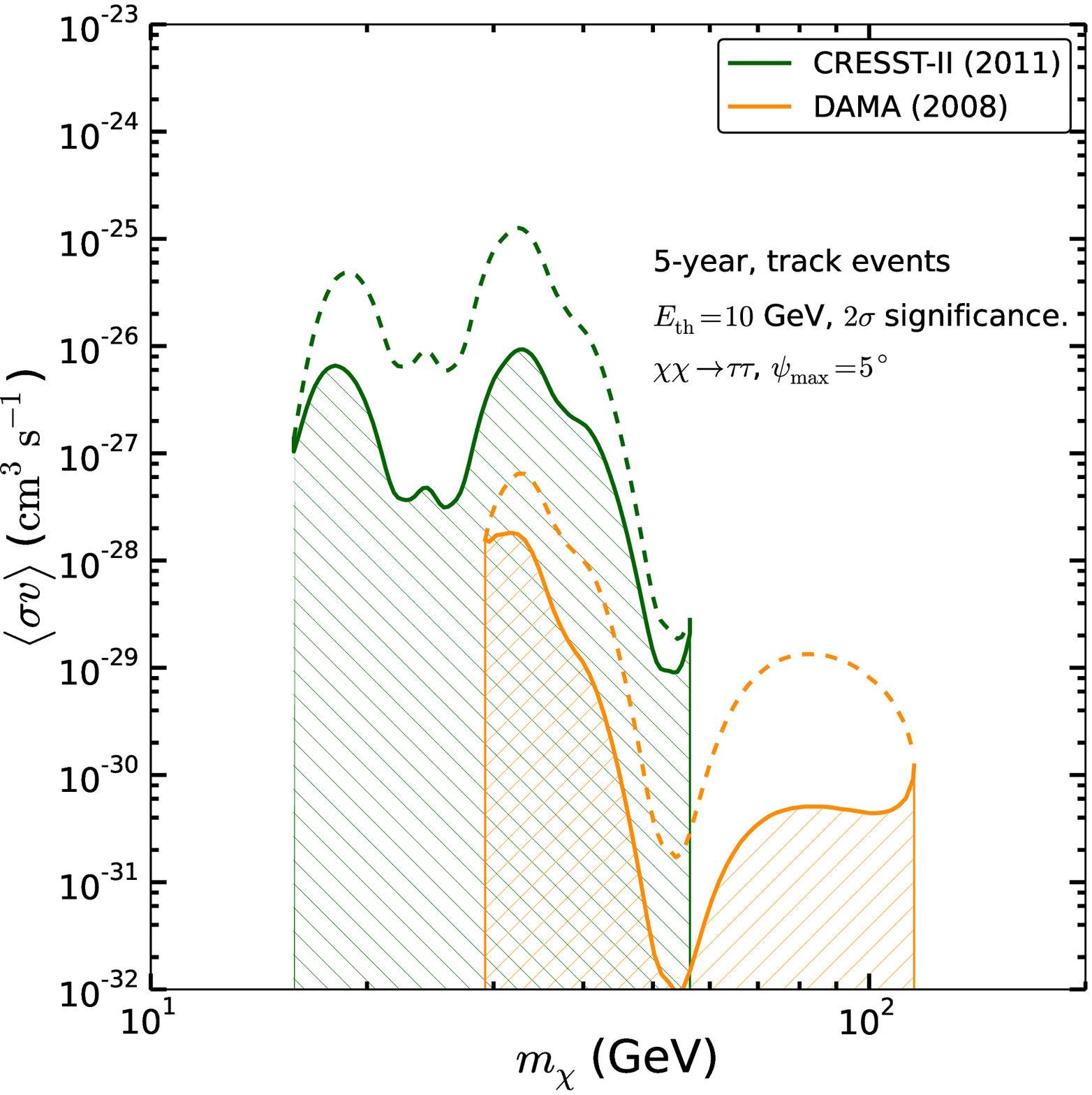}
\caption[]{The IceCube/DeepCore 5 year track-event sensitivity curves in $2\sigma$ significance on the ($m_\chi$,$\sigmav$) 
plane for $E_{\rm{th}}=10~\rm{GeV}$ and $\psi_{\rm{max}}=5^{\circ}$. 
The $\sigsip$ input values are taken from no channeling DAMA and 
CRESST-II shown in Fig.~\ref{fig:mx_sigsip}. The figure in the left panel is for $\chi\chi\to b\bar{b}$ channel while the figure on the right panel is for $\chi\chi\to \tau^+\tau^-$ channel.
The dashed/solid lines correspond to sensitivities obtained with input $\sigsip$ given by the lower/upper boundary of DAMA and CRESST contours.
The allowed regions for $\sigmav$ are indicated by hatched areas. }
\label{fig:mx_sigv_DAMA}
\end{figure} 
%%%%%%%%%%%%%%%%%%%%%%%%%%%%%%%%%%%%%%%%%%%%%%%%%%%%%%%%%%%%%%%%%%%%%%%%%%%%%%%%
From Fig. \ref{fig:sigsip_BF}, we can see that the sensitivity 
is correlated as $\sigsip\sim\sigmav^{-k}$. 
Therefore, if one takes  DAMA and CRESST-II favored regions as input, it is possible to probe $\sigmav$ to a value much smaller than 
$3\times 10^{-26} \rm{cm}^3 s^{-1}$
as shown by  Fig. \ref{fig:mx_sigv_DAMA}.
In this figure, 
it is assumed  that  DM parameter regions are those given by 
DAMA and CRESST-II.  We then present the IceCube/DeepCore  5 year track-event sensitivity 
curves on the ($m_\chi$, $\sigmav$) plane.  We take the boundary of the discovery contour of DAMA and CRESST-II 
as our $\sigsip$ inputs. The upper IceCube/DeepCore sensitivity curves (thin lines) are driven 
by the lower boundary of $\sigsip$ contours of DAMA and CRESST-II, 
while the lower sensitivity curves (thick lines) are driven 
by the upper boundary of $\sigsip$ contours. 
%In some sense, the enclosed contours in Fig. \ref{fig:mx_sigv_DAMA} 
%are showing the uncertainties of $\sigsip$ for our $2\sigma$ sensitivity upper limits.   
We do not use the data of CoGent  
because their favoured DM mass range is below the energy threshold $E_{\rm{th}}=10$ GeV. 
For the same reason, we also ignore the $m_\chi<15$ GeV favoured regions of  
DAMA and \mbox{CRESST-II}. 
We choose the open angle $\psi_{\rm{max}}=5^{\circ}$. 
Because of the large capture rate $C_c$
resulted from the iron resonance 
region, $\sigmav$ for this $m_{\chi}$ range can be probed to values much smaller than the thermal average cross section 
$\sigmav \sim 3\times 10^{-26}$ $\rm{cm}^3 s^{-1}$.

%%%%%%%%%%%%%%%%%%%%%%%%%%%%%%%%%%%%%%%%%%%%%%%%%%%%%%%%%%%%%%%%%%%%%%%%%%%%%%%%
\subsection{$\sigmav$ exclusion limits implied by XENON100 bound on $\sigsip$}

\begin{figure}[ht!]
\centering
\includegraphics[width=0.55\textwidth]{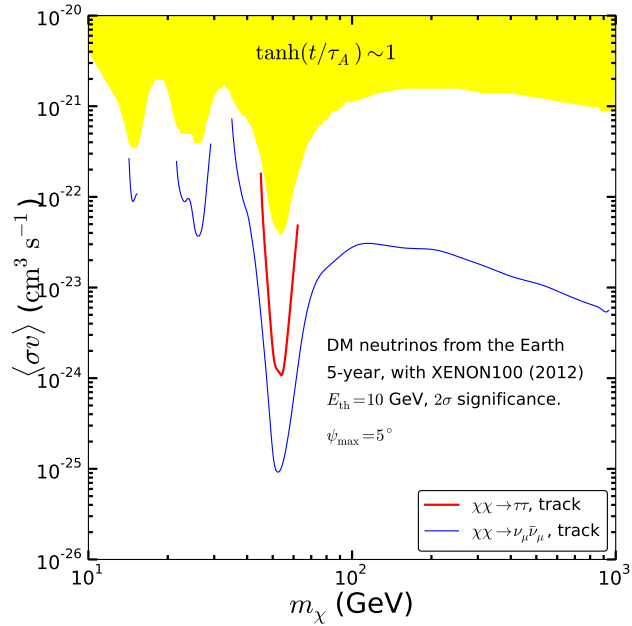}
\caption[]{ The 5-year IceCube/DeepCore $2\sigma$ sensitivities to $\langle\sigma (\chi\chi\to \nu_\mu\bar{\nu}_\mu)\upsilon\rangle$
and $\langle\sigma (\chi\chi\to \tau^+\tau^-)\upsilon \rangle$ 
as functions of $m_{\chi}$ with track events. The $\sigsip$ is taken from XENON100 90\% upper limit. 
The yellow shaded area corresponds to the steady state with $\tanh(\frac{t}{\tau_A})\sim 1$. 
We take $E^{\rm th}=10$ GeV and $\psi_{\rm max}=5^{\circ}$. 
}
\label{fig:mx_sigmav1}
\end{figure}

We can take the XENON100 90\% upper limit as the input $\sigsip$. Let us begin by taking $E^{\rm th}=10$ GeV and consider only track events.
The 5-year IceCube/DeepCore $2\sigma$ sensitivities to $\langle\sigma (\chi\chi\to \nu_\mu\bar{\nu}_\mu)\upsilon\rangle$
and $\langle\sigma (\chi\chi\to \tau^+\tau^-)\upsilon \rangle$ 
as functions of $m_{\chi}$ are presented in Fig.~\ref{fig:mx_sigmav1}. We note that the annihilation channel $\chi\chi\to \tau^+\tau^-$ 
can give rise to track events due to the $\nu_{\tau}\to \nu_{\mu}$ oscillations for lower $E_{\nu}$.  
The results are obtained by considering track events with $\psi_{\rm max}=5^{\circ}$.
The yellow shaded region corresponds to the steady state with $\tanh(\frac{t}{\tau_A})\sim 1$, which is caused by 
a sufficiently large $\sigmav$ when $\sigsip$ is fixed at the current  XENON100 upper limit. 
In this case, the number of DM trapped in the Earth core reaches to the equilibrium value since $\Gamma_A=C_c/2$. 
As a result, the annihilation rate which dictates the neutrino flux is determined entirely by the
capture rate and is independent of $\sigmav$. The latter only determines the number of DM in the equilibrium. Hence 
the measurement of neutrino flux in the steady state can only determine $\sigsip$. 
%%%%%%%%%%%%%%%%%%%%%%%%%   F   I   G   U   R   E   %%%%%%%%%%%%%%%%%%%%%%%%%%%%
%%%%%%%%%%%%%%%%%%%%%%%%%%%%%%%%%%%%%%%%%%%%%%%%%%%%%%%%%%%%%%%%%%%%%%%%%%%%%%%%
 
We can see the strongest limit comes from the iron resonance region $m_\chi\sim 50$ GeV where the capture rate $C_c$ peaks.
On the other hand, the weakest bound 
of $\sigmav$ occurs at the
lowest point of XENON100 $\sigsip$ upper limit located at $m_\chi\sim 100-200$ GeV. 
Moreover, the curves are broken near the boundary of the yellowed region. Hence, for each annihilation channel,
%for specific annihilation channels,  
there exists a range of $m_\chi$ 
where the $2\sigma$ significance curve disappears. This means that 
the DM event number in this $m_\chi$ range cannot reach the $2\sigma$ significance before $\sigmav$ reaches to the steady state
$\tanh(\frac{t}{\tau_A})\sim 1$. 
One should bear in mind that this yellowed region varies with the input $\sigsip$.

%%%%%%%%%%%%%%%%%%%%%%%%%   F   I   G   U   R   E   %%%%%%%%%%%%%%%%%%%%%%%%%%%%
\begin{figure}[ht!]
\centering
\includegraphics[width=0.49\textwidth]{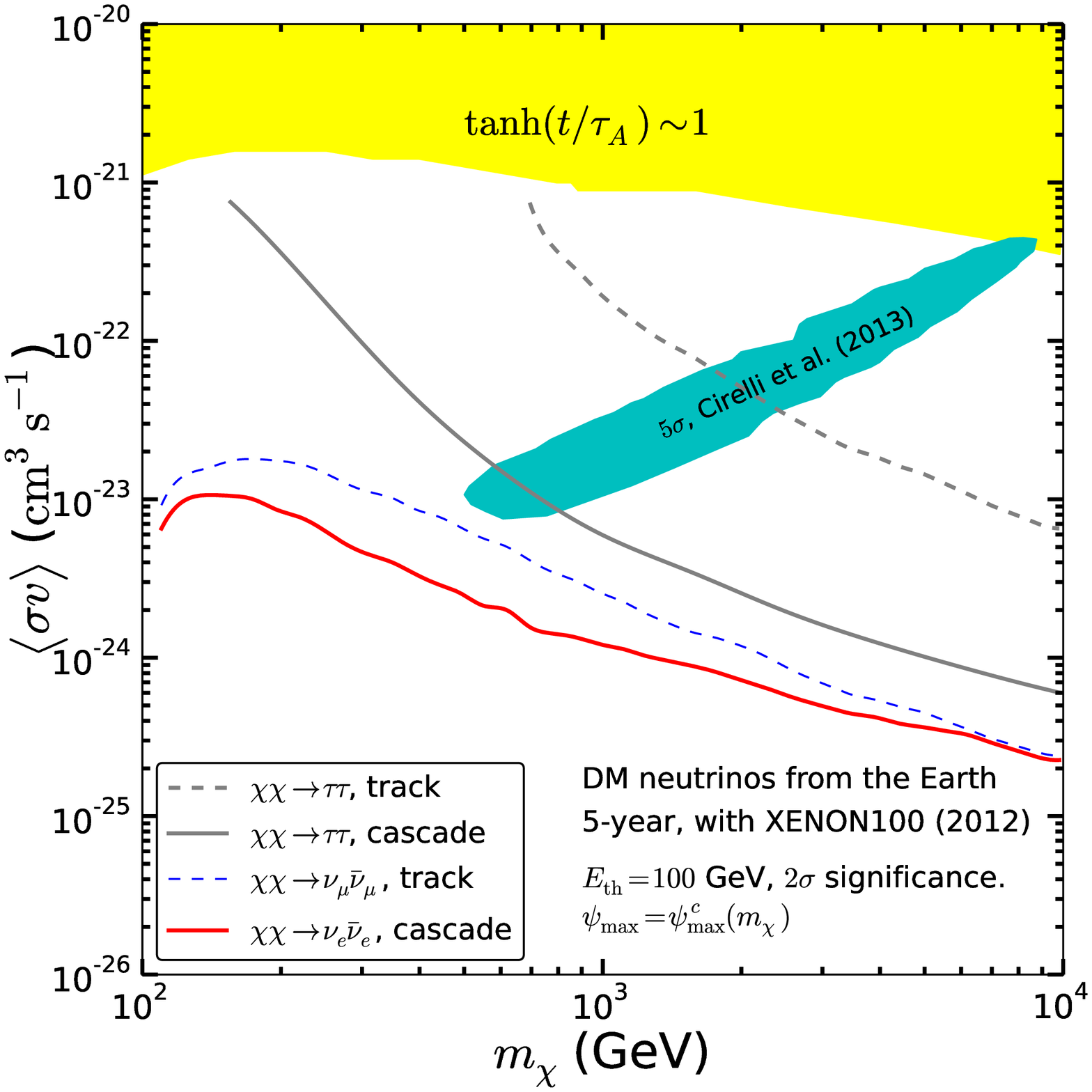}
\includegraphics[width=0.49\textwidth]{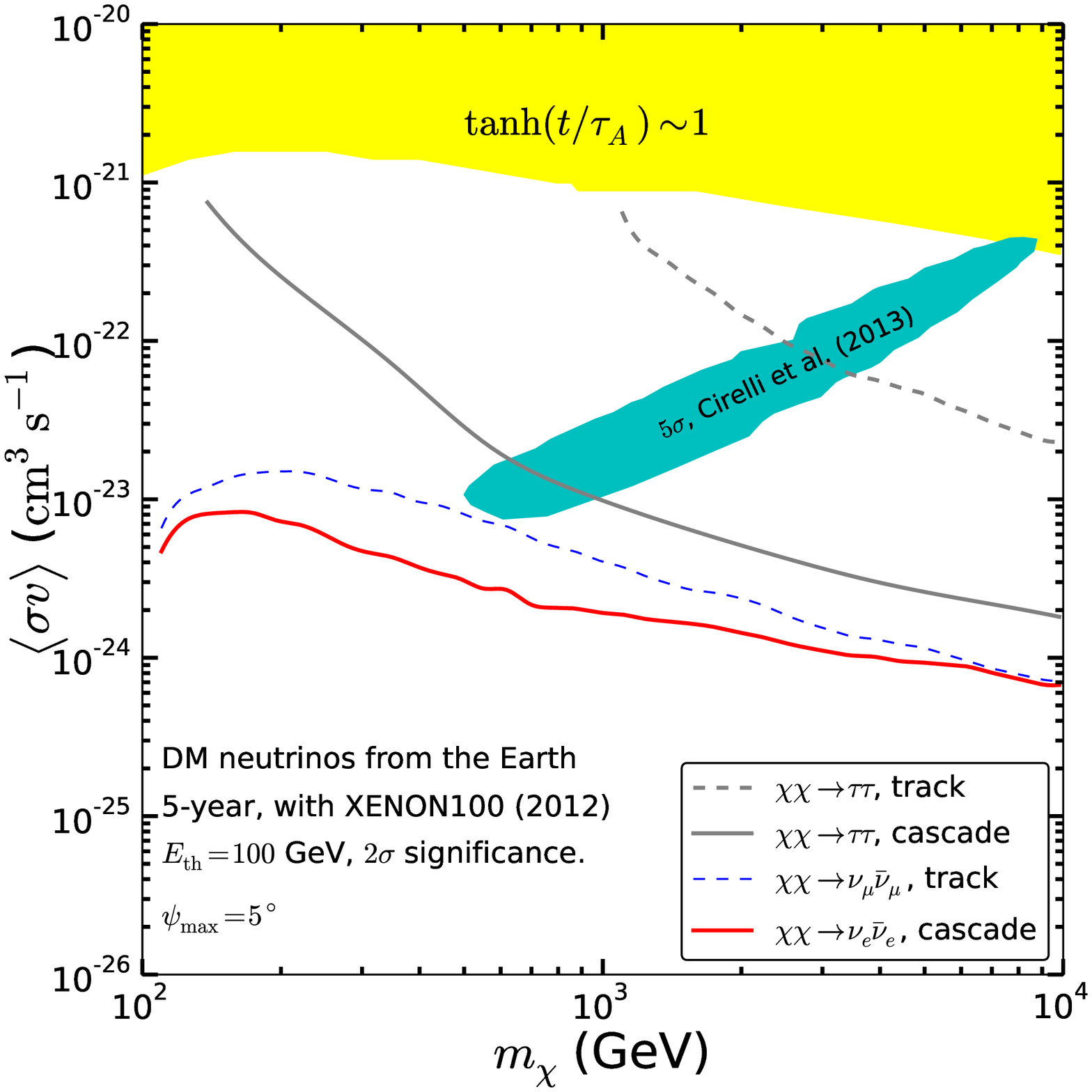}\\
\includegraphics[width=0.49\textwidth]{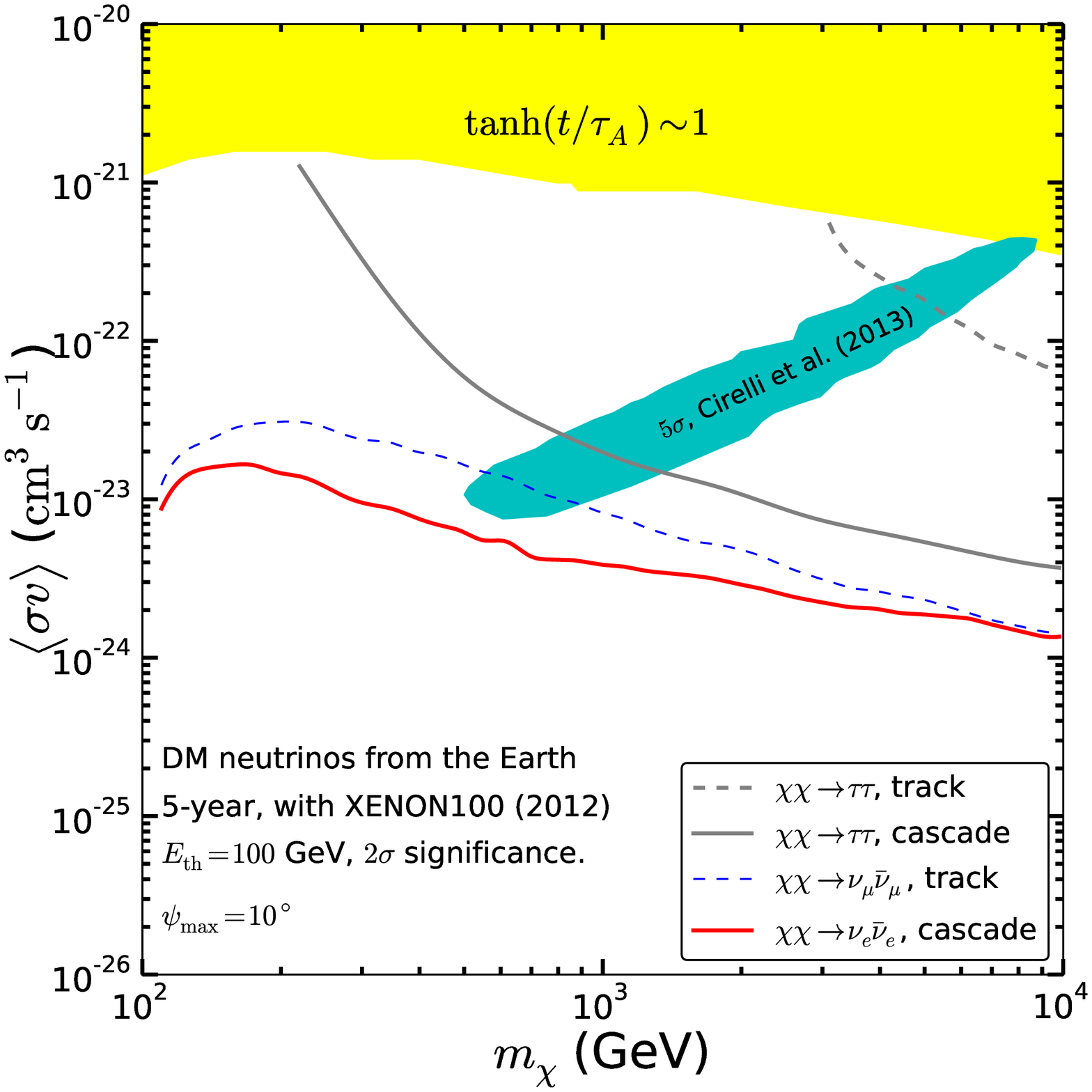}
\includegraphics[width=0.49\textwidth]{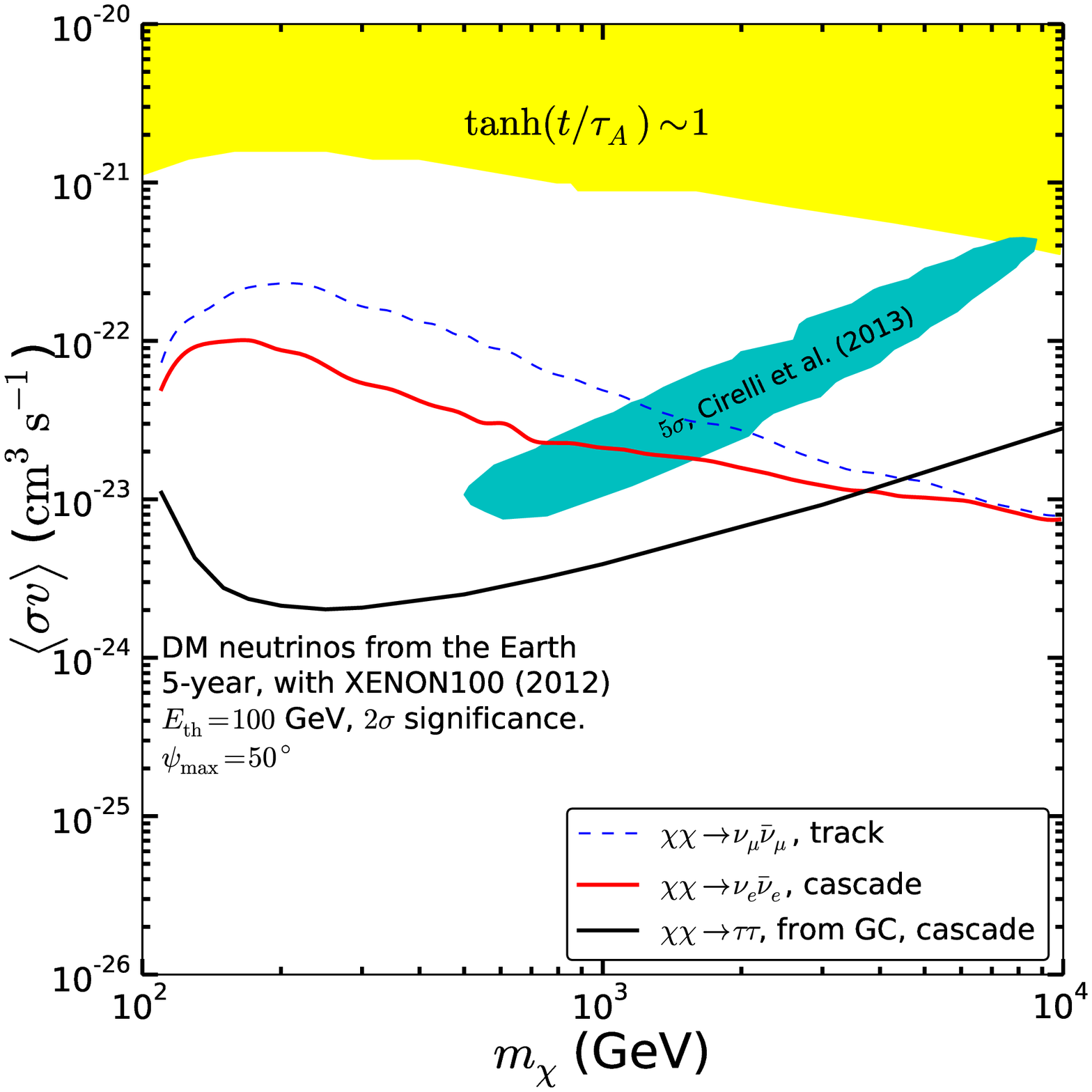}\\
\caption[]{The 5-year IceCube/DeepCore $2\sigma$ sensitivities to $\sigmav$ of various channels 
as functions of $m_{\chi}$ with track and cascade events. The threshold energy is taken to be $100$ GeV
and the $m_\chi$ range is extended to 10 TeV. The cyan contours refer to $5\sigma$ 
confidence region of PAMELA and AMS02 combined analysis, for $\chi\chi\to \tau^+\tau^-$ channel,
taken from Ref.~\cite{Cirelli:2008pk}. The black solid line is the 5 year sensitivity upper limit of $\langle \sigma (\chi\chi\to \tau^+\tau^-)\upsilon \rangle$ 
in $2\sigma$ significance obtained from the search of neutrino cascade events from the galactic center with $\psi_{\rm{max}}=50^{\circ}$ relative to the direction of galactic center.
}
\label{fig:mx_sigmav2}
\end{figure} 
%%%%%%%%%%%%%%%%%%%%%%%%%%%%%%%%%%%%%%%%%%%%%%%%%%%%%%%%%%%%%%%%%%%%%%%%%%%%%%%% 
We next take $E^{\rm th}=100$ GeV and consider both track and cascade events. We present in  Fig.~\ref{fig:mx_sigmav2} 
the IceCube/DeepCore sensitivities to $\sigmav$ of various channels 
as functions of $m_{\chi}$.
The result in the upper left panel is obtained by taking $\psi_{\rm max}=\psi^{\rm c}_{\rm max}(m_\chi)$, while  
$\psi_{\rm{max}}$ is taken to be $5^{\circ}$, $10^{\circ}$ and $50^{\circ}$ 
for results in the upper right, lower left and lower right panels, respectively. 
For $\chi\chi\to \tau^+\tau^-$ channel, we also plot the $5\sigma$ confidence region 
favoured by PAMELA and AMS02 positron fraction data~\cite{Cirelli:2008pk} for comparison. 
One can see that the search for $\chi\chi\to \tau^+\tau^-$ cascade events can probe almost all the $5\sigma$ favoured region by PAMELA and AMS02 
with $\psi_{\rm max}=5^{\circ}$. 
For comparison, we also estimate the IceCube/DeepCore sensitivity to the 
DM annihilation cross section in the galactic halo with a 100 GeV threshold energy by 
using the method of Ref.~\cite{Lee:2012pz} and the energy 
dependent effective volume $V_{\textrm{eff}}(E)$ \cite{Collaboration:2011ym}. 
The 5 year sensitivity upper limit of $\langle \sigma (\chi\chi\to \tau^+\tau^-)\upsilon \rangle$ 
obtained from searching for neutrino cascade events from the galactic center with $\psi_{\rm{max}}=50^{\circ}$ relative to the direction of galactic center
is also plotted. We note that this sensitivity is independent of $\sigsip$, unlike 
the search for DM annihilations in the Earth core. 
One can see that the search for galactic DM annihilations by IceCube/DeepCore can probe the entire $5\sigma$ confidence region favoured by PAMELA and AMS02 
in 5 years of running.   

\begin{figure}[ht!]
\centering
\includegraphics[width=0.55\textwidth]{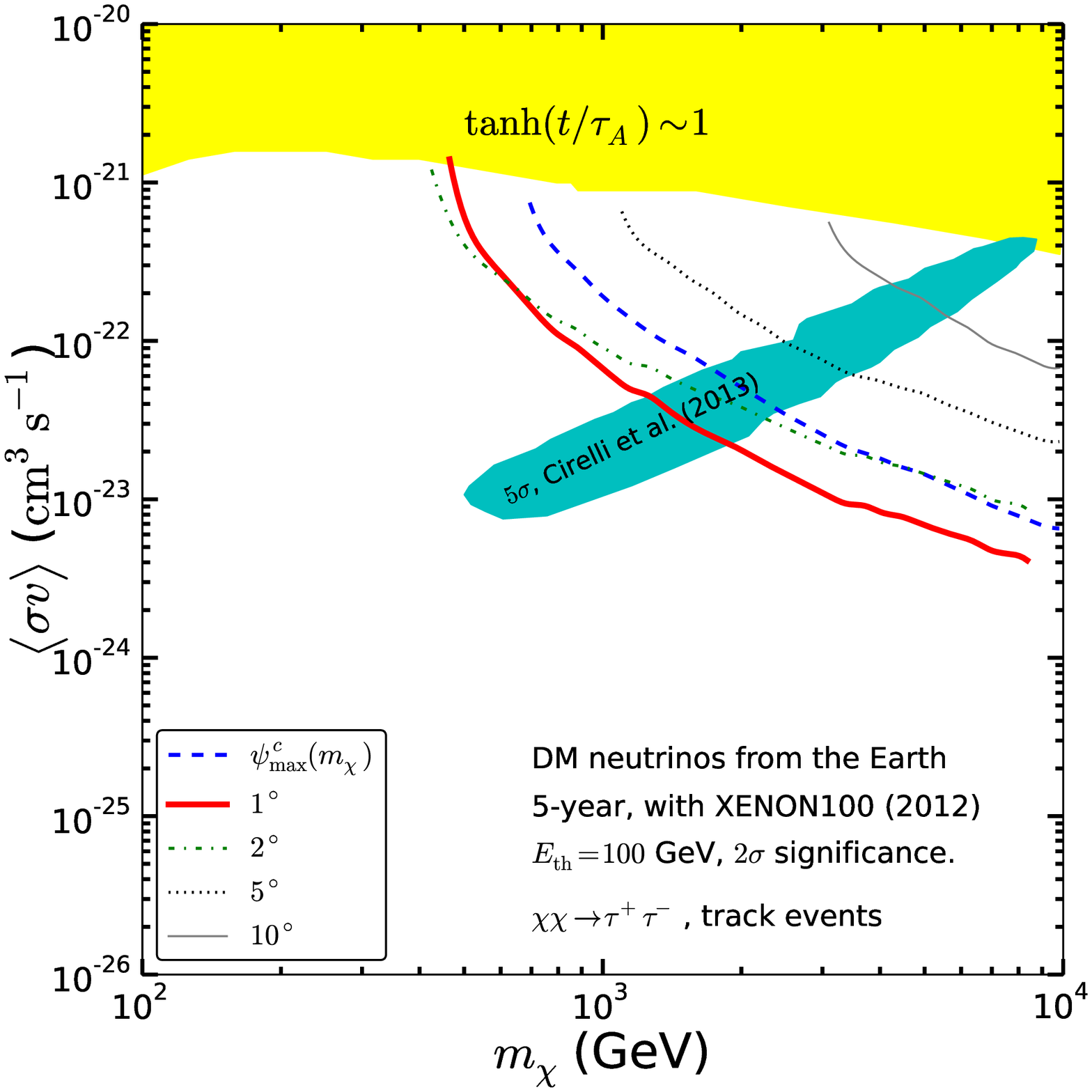}
\caption[]{The IceCube/DeepCore sensitivities to $\langle \sigma(\chi\chi\to \tau^+\tau^-)\upsilon\rangle$ with track events 
for several different open angles.
}
\label{fig:mx_sigv_tautau}
\end{figure}

It is interesting to note that $\chi\chi\to \tau^+\tau^-$ at the current energy range also produces track events since tau lepton can decay into 
muon neutrinos. 
In Fig. \ref{fig:mx_sigv_tautau}, we summarize the sensitivities 
with track events for $\tau^+\tau^-$ channel with different open angles, 
$\psi_{\rm max}=1^{\circ}$ (red solid), $2^{\circ}$ (green dashed-dot), $5^{\circ}$ (black dot), $10^{\circ}$ (grey solid), 
and $\psi_{\rm max}=\psi^{\rm c}_{\rm{max}}(m_\chi)$ (blue dashed).   
We note that the sensitivity curves for $\psi_{\rm max}=1^{\circ}$ and $2^{\circ}$ cross at $m_\chi\sim 600$ GeV. 
In other words, for $m_\chi< 600$ GeV, the numerator in Eq.~(\ref{eq:sensitivity}) increases faster than the denominator as $\psi_{\rm max}$ increases from $1^{\circ}$ to $2^{\circ}$.
(see the left panel of Fig.~\ref{fig:Nevents} ).   
Moreover, we note that the sensitivity to $\sigmav$ can change by more than one order of magnitude 
as the open angle $\psi_{\rm max}$ increases from $1^{\circ}$ to $10^{\circ}$.    
Finally, with $\psi_{\rm max}=1^{\circ}$, IceCube 5 year data  can probe the 
PAMELA and AMS02 positron favoured region for $m_\chi\gtrsim 2$ TeV.  

%%%%%%%%%%%%%%%%%%%%%%%%%   F   I   G   U   R   E   %%%%%%%%%%%%%%%%%%%%%%%%%%%%
%%%%%%%%%%%%%%%%%%%%%%%%%%%%%%%%%%%%%%%%%%%%%%%%%%%%%%%%%%%%%%%%%%%%%%%%%%%%%%%% 

%%%%%%%%%%%%%%%%%%%%%%%%%%%%%%%%%%%%%%%%%%%%%%%%%%%%%%%%%%%%%%%%%%%%%%%%%%%%%%%%
\subsection{ Pessimistic scenario by assuming  non-detection of XENON1T (2017)}

%%%%%%%%%%%%%%%%%%%%%%%%%   F   I   G   U   R   E   %%%%%%%%%%%%%%%%%%%%%%%%%%%%
\begin{figure}[ht!]
\centering
\includegraphics[width=0.55\textwidth]{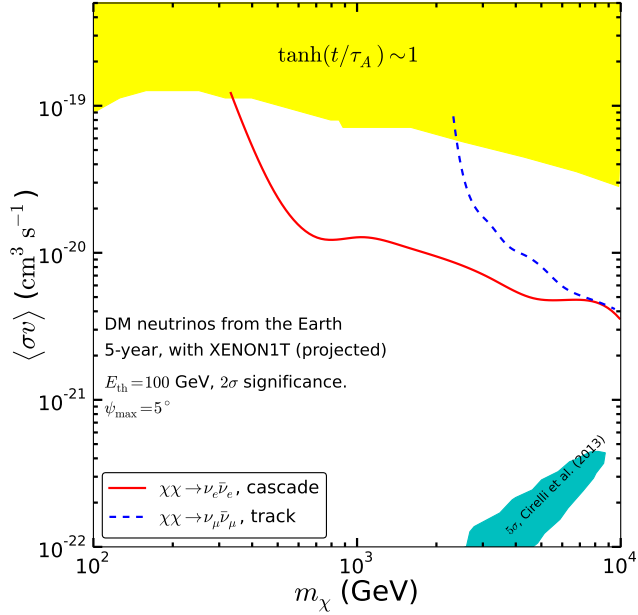}\\
\caption[]{ 
The IceCube/DeepCore 5 year sensitivity with $2\sigma$ significance on the
($m_{\chi}$, $\sigmav$) plane. The $\sigsip$ input is taken from 
XENON1T sensitivity curve. 
We take $\psi_{\rm max}=5^{\circ}$. The threshold energy is taken to be $100$ GeV. The red solid line 
and blue dashed line are IceCube sensitivities to $\chi\chi\to \nu_e\bar{\nu}_e$ cascade events and $\chi\chi\to \nu_\mu\bar{\nu}_\mu$ 
track events, respectively.
}
\label{fig:mx_sigv_X1T}
\end{figure} 
%%%%%%%%%%%%%%%%%%%%%%%%%%%%%%%%%%%%%%%%%%%%%%%%%%%%%%%%%%%%%%%%%%%%%%%%%%%%%%%% 

Finally we discuss a pessimistic scenario that DM is not detected
by XENON1T (a future ton-size DM detector).
In Fig.~\ref{fig:mx_sigv_X1T}, instead of using the current XENON100 limit as the input for $\sigsip$, 
we compute the IceCube 5-year $2\sigma$ sensitivity upper limit 
with the projected $\sigsip$ limit from XENON1T \cite{Aprile:2012zx} as the input. 
We set the threshold energy at 100 GeV 
and present our result for $m_{\chi}$ up to 10 TeV. Even if XENON1T $\sigsip$ sensitivity limits
at larger $m_{\chi}$ are weaker, only those $\sigmav$ arising from monochromatic annihilation channels 
can be probed by IceCube, i.e., by observing cascade events from $\chi\chi\to \nu_e\bar{\nu}_e$ and  
$\chi\chi\to \nu_\tau\bar{\nu}_\tau$ channels (not shown on the figure), and by observing track events from $\chi\chi\to \nu_\mu\bar{\nu}_\mu$ channel.
%Therefore, if DM would not be detected by XENON1T, 
%only those $\sigmav$ from monochromatic annihilation channels can be constrained by IceCube for $m_{\chi}\lesssim 10$ TeV. 
However this upper limit is not stringent since the strongest bound for $\sigmav$ in this case is roughly $10^{-22}$ cm$^3$s$^{-1}$.

\section{SUMMARY AND CONCLUSIONS}
In this paper, we study the neutrino signature arising from DM annihilations inside 
the Earth core. Applying IceCube/DeepCore effective areas, 
we have computed IceCube/DeepCore 5-year sensitivities in $2\sigma$ significance with  track and cascade events. 
From the slope of sensitivity curves in Fig.~\ref{fig:sigsip_BF}, the neutrino event rate is more sensitive to $\sigsip$ than $\sigmav$. 
To illustrate the impact of $\sigsip$ on the neutrino event rate, 
we have focused on three different scenarios according to different input $\sigsip$. 
Hence our results can be divided by three categories.    

\underline{\textbf{Implications from $\sigsip$ favored by of DAMA and CRESST II:}}\\

We have compared the 5-year full IceCube/DeepCore sensitivity to $\sigsip$ derived from the search for DM annihilations in the Earth core 
with the recent limit on $\sigsip$ by IceCube 79-string detector search for DM annihilations in the Sun. 
We found that the small DM mass region, $m_\chi\lesssim 100$ GeV, can be better probed 
by detecting DM annihilations in the Earth core.  
By fixing $\sigmav=3\times 10^{-26} \rm{cm}^3 s^{-1}$, we found  
that our $\tau^+\tau^-$ track event result can probe 
the entire DAMA allowed region and most of the CRESST-II allowed region. 
If one takes the large $\sigsip$ favored by DAMA and CRESST-II as input, 
a rather low $\sigmav$ is sufficient 
for IceCube/DeepCore to achieve  $2\sigma$ detection significance in 5 years on DM annihilations in the Earth core. 
It will be quite challenge for other indirect detection experiments to achieve such a sensitivity to 
$\sigmav$ 
in the near future. 
%On the other hand, the IC86 might test these lower $\sigmav$ regions 
%by the detection of neutrino from the Earth.

\underline{\textbf{Implication from $\sigsip$ bound set by XENON100:}}\\

We have also considered the scenario of taking the current XENON100 bound as our input $\sigsip$.
We have discussed the implications by taking $E^{\rm th}$ as $10$ GeV and $100$ GeV, respectively.  
In the former case, we study the IceCube/DeepCore sensitivities for $m_{\chi}$ up to $1$ TeV and
consider only track events. In the latter case,  the IceCube/DeepCore sensitivities are studied for $m_{\chi}$ up to $10$ TeV 
with both track and cascade events.

For $E_{\rm th}=10$ GeV, we found that 
the strongest limit of $\sigmav$ comes from 
the iron resonance region $m_\chi\sim 50$ GeV. 
Among all the DM annihilation channels,
the most stringent limit with track events arises  from $\chi\chi\to \nu_\mu\bar{\nu}_\mu$ annihilation. 

For $E_{\rm th}=100$ GeV where both track and cascade events are considered, 
the most stringent limit of $\sigmav$ is from $\chi\chi\to \nu_e\bar{\nu}_e$ cascade events. 
Moreover, we have compared the IceCube/DeepCore sensitivities on the $(m_\chi, \sigmav)$ plane with the parameter range favored by PAMELA 
and AMS02 data.  
We found that both track and cascade events in $\chi\chi\to \tau^+\tau^-$ annihilation channel can  
test the PAMELA and AMS02 favored parameter space.
With $\psi_{\rm max}=1^{\circ}$, the search for DM induced neutrino track events from the Earth core  
can rule out the PAMELA and AMS02 favoured parameter region at $m_\chi\gtrsim 2$ TeV.

\underline{\textbf{Implication of future XENON1T sensitivity:}}\\
 
We finally discussed the pessimistic scenario that DM is not discovered by  
the future XENON1T. With an input $\sigsip$ given by XENON1T sensitivity, we again discuss the implication
on neutrino search.     
 
With $E^{\rm th}=100$ GeV and $10^2<m_\chi/{\rm GeV}<10^4$, we found that  
only those $\sigmav$ arising from monochromatic annihilation channels, $\chi\chi\to \nu\bar{\nu}$, 
can be probed by IceCube in 5 years of data taking.
However the expected bound on $\sigmav$ by IceCube is disfavored by the current AMS-02 positron flux result.

\section*{Acknowledgments}
 F.F.L. is supported by the grant from Research and Development Office, National Chiao-Tung University, G.L.L. is supported by National Science Council
of Taiwan under Grant No. 99-2112-M-009-005-MY3 and National Center for Theoretical Sciences (NCTS), Taiwan,
and Y.S.T. is funded in part by the Welcome Programme of the Foundation for Polish Science. 
Y.S.T. also likes to thank NCTS for hospitality during his visit.

\end{document}